\documentclass[12pt,pre,aps,preprint]{revtex4}

\usepackage{color}
\usepackage{graphics}
\usepackage{graphicx}
\usepackage{amsfonts}
\usepackage{amsmath}
\usepackage{float}
\usepackage{epsfig}

\begin{document}
\title{Effect of Dimensionality on the Percolation Thresholds of Various $d$-Dimensional
 Lattices}

\author{S. Torquato}

\email{torquato@electron.princeton.edu}

\affiliation{\emph{Department of Chemistry, Department of Physics,
Princeton Institute for the Science and Technology of
Materials, and Program in Applied and Computational Mathematics}, \emph{Princeton University},
Princeton NJ 08544}

\author{Y. Jiao}

\email{yjiao@princeton.edu}

\affiliation{\emph{Princeton Institute for the Science and Technology of
Materials}, \emph{Princeton University},
Princeton NJ 08544}

\begin{abstract}
We show analytically that the $[0,1]$, $[1,1]$ and $[2,1]$   Pad{\'e } approximants
of the mean cluster number $S(p)$ for site and bond percolation on general
$d$-dimensional lattices are upper bounds on this quantity in any Euclidean dimension $d$,
where $p$ is the occupation probability. These results lead to certain lower bounds 
on the percolation threshold $p_c$ that become
progressively tighter as $d$ increases and asymptotically exact as $d$ becomes large.
These lower-bound estimates depend on the structure of the $d$-dimensional lattice 
and whether site or bond percolation is being considered.
We obtain explicit bounds on $p_c$ for both site and bond percolation
on five different lattices:
$d$-dimensional generalizations of the simple-cubic, body-centered-cubic
and face-centered-cubic Bravais lattices as well as the $d$-dimensional
generalizations of the diamond and kagom{\'e } (or pyrochlore) non-Bravais
lattices. These analytical estimates are used to assess available simulation results
across dimensions (up through $d=13$ in some cases). It is noteworthy that the tightest lower bound provides
reasonable estimates of $p_c$ in relatively low dimensions and becomes increasingly
accurate as $d$ grows. We also derive high-dimensional asymptotic
expansions for $p_c$ for the ten percolation problems and compare
them to the Bethe-lattice approximation. Finally, we remark on the radius of convergence
of the series expansion of $S$ in powers of $p$ as the dimension grows.
\end{abstract}


\maketitle
\section{Introduction}

There has been a longstanding interest to understand the effect of dimensionality on the
structure and bulk properties of models of condensed phases of matter, especially
lattice models \cite{Fi61a,Fi64,Es72,Ga76,Ga78,Sah83,Bi92}. More recently,
the high-dimensional behavior of interacting many-particle systems
has received considerable attention and led to insights into low-dimensional
systems. This includes studies of models of liquids and glasses \cite{Fr99,Pa00,Sk06,To06a,Ro07,Me09,Pa10,Lu10,To10c}, hyperuniformity of many-particle configurations and their local density fluctuations \cite{To03a,Za09}, covering and quantizer problems \cite{Co93} and their relationships to classical ground states \cite{To10d}, densest sphere packings  \cite{Co03,To06b},
and Coulombic systems \cite{To08c}.
The preponderance of studies aimed at elucidating the dependence of dimensionality
across all dimensions have been carried out for Ising-spin and lattice-percolation models;
see, among the multitude of such investigations, Refs.  \onlinecite{Fi64,Es72,Ga76,Ga78,Sah83,Bi92}.
Virtually all of this work has been carried out on the $d$-dimensional hypercubic lattice $\mathbb{Z}^d$.
The present paper is concerned with the prediction of Bernoulli nearest-neighbor
site and bond percolation thresholds on general $d$-dimensional lattices in Euclidean space $\mathbb{R}^d$.

While it is well-known that critical exponents first take on their mean-field dimension-independent
values when $d=6$, independent of the lattice, the percolation thresholds $p_c$ generally
depend on structure of the lattice and are believed to achieve their mean-field
values only in the limit of infinite dimension \cite{Fi61a}.
Whereas thresholds are known exactly for only a few lattices in two dimensions \cite{Sy64},
there are no such exact results for $d\ge 3$ for finite $d$. Thus, most studies
of the determination of lattice thresholds in any finite dimension
have relied on numerical methods or approximate theoretical techniques \cite{Do61,Ar79,Sah83,Stell85,Given91,St92,Sa94,Lo98,Lor98,Va98a,Va98b,Ne00,Lor00,Gr03,Sk09,Zi09,Di10}.

It has recently been shown that the  $[0,1]$, $[1,1]$ and $[2,1]$ Pad{\'e } approximants
of the density-dependent mean cluster number $S$ for prototypical $d$-dimensional continuum
percolation models provide lower bounds on the corresponding thresholds \cite{To12a}.
Specifically, these results apply to overlapping (Poisson distributed) hyperspheres
as well as hyperparticles of nonspherical shapes with some specified orientational
distribution function. The sharpness of these bounds showed that previous simulations
for the thresholds were inaccurate in higher dimensions, which then led to studies
that reported improved estimates for the thresholds
of overlapping hyperspheres \cite{To12c} as well as for overlapping hyperparticles with a variety of specific
shapes \cite{To13a} that apply in any dimension $d$.

Using the same techniques as was employed in Ref. \onlinecite{To12a}, we obtain analogous lower bounds on $p_c$
for site and bond percolation for general $d$-dimensional lattices in $\mathbb{R}^d$.
We demonstrate that these general lower bounds become
progressively tighter as $d$ increases and exact asymptotically as $d$ becomes large.
Employing these general results, we derive explicit expressions for lower bounds on $p_c$ for five distinct lattices:
$d$-dimensional generalizations of the simple-cubic, body-centered-cubic
and face-centered-cubic Bravais lattices as well as the $d$-dimensional
generalizations of the diamond and kagom{\'e } (or pyrochlore) non-Bravais
lattices. Our analytical lower-bound estimates of these ten different percolation problems  are then employed to assess available simulation results
across dimensions (up through $d=13$ in some cases). We show that the tightest lower bound provides
reasonable estimates of $p_c$ in relatively low dimensions and becomes increasingly
accurate as $d$ grows. Our investigation also sheds light on the radius of convergence of the series
expansion of the mean cluster number $S(p)$ in powers of the occupation probability $p$
across dimensions.

The rest of the paper is organized as follows: We provide fundamental definitions in Sec. II 
and derive lower bounds on the percolation threshold $p_c$ in Sec. III. In Sec. IV, we describe 
the $d$-dimensional lattices that will be considered here as well as obtain series expansions 
of $S(p)$ and asymptotic expansions of the lower bounds on $p_c$. In Sec. V, we explicitly evaluate 
the bounds on $p_c$ across dimensions and compare them to available simulation results. We close with
concluding remarks and discussion in Sec. V1.

\section{Definitions and Preliminaries}
\label{defs}

\subsection{Bravais and Non-Bravais Lattices}

A $d$-dimensional Bravais lattice in $\mathbb{R}^d$ is the set of
points defined by integer linear combinations of a set of basis
vectors, i.e., each site is specified by the {\it lattice vector}
\begin{equation}
{\bf p}= n_1 {\bf a}_1+ n_2 {\bf a}_2+ \cdots + n_{d-1} {\bf a}_{d-1}+n_d {\bf a}_d,
\end{equation}
where ${\bf a}_i$ are the basis vectors for the fundamental cell,
which contains just {\it one} point, and $n_i$ spans all the
integers for $i=1,2,\cdots d$. Every Bravais lattice has a {\it
dual} or {\it reciprocal} Bravais lattice in which the sites of
the lattice are specified by the dual (reciprocal) lattice vector
${\bf q}$ such that ${\bf q}\cdot {\bf p}=2\pi m$, where $m= \pm
1, \pm 2, \pm 3 \cdots$; see Conway and Sloane ~\cite{Co93} for
additional details. The concept of a Bravais lattice can be
naturally generalized to include multiple points within the
fundamental cell, defining a periodic crystal or non-Bravais
lattice. Specifically, a non-Bravais lattice consists of the union
of a Bravais lattice with one or more translates of itself; it can
therefore be defined by specifying the lattice vectors for the
Bravais lattice along with a set of translate vectors that define
the {\it basis} (number of points per fundamental cell).

\subsection{Connectedness Criterion}

Consider a $d$-dimensional lattice $\Lambda$ in $\mathbb{R}^d$  in which each site is occupied
with probability $p$ in the case of site percolation or in which each bond
is occupied with probability $p$ in the case of bond percolation. The lattice
$\Lambda$ can either be a Bravais or non-Bravais lattice.
We consider Bernoulli percolation with a nearest-neighbor connectivity criterion
for either site or bond percolation on $\Lambda$ in which the
{\it coordination number} $z_{\Lambda}$ is the number of nearest neighbors.  The following indicator
function defines this connectivity criterion:
\begin{equation}
f({\bf r}_{ij}) =\Bigg\{{1 \; \mbox{ if sites (or bonds) $i$ and $j$ are occupied nearest neighbors},\atop{0, \; \mbox{otherwise}}}
\label{mayer}
\end{equation}
where ${\bf r}_{ij}$ is the displacement vector between sites (or bonds) $i$ and $j$.
In the case of site percolation,
\begin{equation}
\sum_{j=1} f({\bf r}_{1j})= z_s =z_{\Lambda}
\label{zs}
\end{equation}
where $z_{\Lambda}$ is the coordination number for the lattice $\Lambda$.
In the case of bond percolation,
\begin{equation}
\sum_{j=1} f({\bf r}_{1j})= z_b = 2(z_{\Lambda}-1)=2(z_s-1),
\label{zb}
\end{equation}
where it is to be noted that generally $z_b > z_s$ for any $d \ge 2$.

\subsection{Connectedness Functions}

The {\em mean cluster number} (or mean cluster size) $S$ is the average number of sites
(bonds) in the cluster containing a randomly chosen occupied site (bond). The {\em pair-connectedness function}
$P_{2}({\bf r})$ is defined such that
$p^2 P_{2}({\bf r})$ gives the probability that a site (center of a bond) at the origin
and a site (bond center) $j$ located at position ${\bf r}$ are both occupied
and belong to the same cluster.
Essam showed that the mean cluster number   is related to
a sum over the pair-connectedness function \cite{Es72}:
\begin{equation}
S=1+p\,\sum_{\bf r}P_{2}({\bf r}).
\label{S1}
\end{equation}
This relation can be  equivalently expressed in terms of the Fourier transform ${\tilde P}({\bf k})$ of $P({\bf r})$:
\begin{equation}
S=1+p\, {\tilde P}({\bf k}={\bf 0}).
\label{S2}
\end{equation}
Using the Ornstein-Zernike equation \cite{Co77} that defines the direct connectedness function
$C({\bf r})$:
\begin{equation}
{\tilde P}({\bf k})={\tilde C}({\bf k})+ p\, {\tilde C}({\bf k}){\tilde P}({\bf k}),
\label{OZ1}
\end{equation}
where ${\tilde C}({\bf k})$ is the Fourier transform  of $C({\bf r})$, we also can express
the mean cluster number as follows:
\begin{equation}
S=[1-p\, {\tilde C}(0)]^{-1}.
\label{S3}
\end{equation}
Since $P({\bf r})$ becomes long-ranged (i.e., decays to zero for large $r$ slower than $1/r^d$),
$S$ diverges in the limit $p \rightarrow p_c^{-}$, and hence we have from (\ref{S3}) that
the percolation threshold is given by
\begin{equation}
p_c = [{\tilde C}(0)]^{-1}.
\end{equation}
It is instructive to note that the real-space equation corresponding to relation (\ref{OZ1}) is
\begin{equation}
P({\bf r}_{12}) = C({\bf r}_{12}) + p\; \sum_{j=1} C({\bf r}_{1j})P({\bf r}_{2j}).
\end{equation}
The sum operation here is the analog of the convolution integral in $\mathbb{R}^d$.

It is believed that $S$ obeys the power law
\begin{equation}
S \propto (p_c - p)^{-\gamma}, \qquad p \to p_c^-
\label{Sasymp}
\end{equation}
in the immediate vicinity of the percolation threshold. In this
expression, $\gamma$ is a universal exponent for a large class of
lattice and continuum percolation models in dimension $d$,
including not only Bernoulli lattice and spatially uncorrelated
continuum models, but correlated continuum systems
\cite{St92,Sa94,To02a}. For example, $\gamma=43/18$ for $d=2$ and
$\gamma=1.8$ for $d=3$. It is believed that when $d\ge d_c=6$,
where $d_c$ is the ``critical'' dimension, the lattice- and
continuum-percolation exponents take on their
dimension-independent {\it mean-field} values,
\cite{St92,Sa94,To02a} which means in the case of (\ref{Sasymp})
that $\gamma=1$. These mean-field values are obtainable exactly
from percolation on an infinite tree, such as the Bethe lattice
for which Fisher and Essam \cite{Fi61a} showed that the threshold
is given by
\begin{equation}
p_c = \frac{1}{z_{\Lambda}-1}.
\label{B1}
\end{equation}
The dimensionality of the Bethe lattice is effectively infinite and therefore
it is generally assumed that $p_c$ for (periodic) lattices approach  the Bethe-lattice approximation (\ref{B1})
in the limit $d \rightarrow \infty$. 
We will see in Sec. \ref{Exact} that this assumption is generally not exactly true. 
Note that for the large class of periodic lattices in which the coordination
number $z_{\Lambda}$ grows monotonically with $d$, the high-dimensional Bethe approximation becomes
\begin{equation}
p_c \sim \frac{1}{z_{\Lambda}} \qquad (d \rightarrow \infty).
\label{B2}
\end{equation}

\subsection{Cluster Statistics }

A $k$-mer is a cluster that contains $k$ sites or bonds. The {\em cluster-size distribution}
$n_k$ is the average number of $k$-mers per site (bond). Thus, the probability
that an arbitrary site (bond) is part of a $k$-mer is $kn_k$, and hence
\begin{equation}
\sum_{k=1}^{\infty} kn_k=p, \qquad p<p_c.
\label{sum1}
\end{equation}
Since the quantity $kn_k/\Sigma_{k}kn_k$ is the
probability that the cluster to which an arbitrary occupied site (bond) belongs contains
exactly $k$ sites (bonds), the mean cluster number $S$ can be alternatively expressed as
\begin{equation}
S=\frac{\displaystyle\sum_{k=1}^\infty k^2 n_k}{\displaystyle\sum_{k=1}^\infty k n_k}, \qquad p<p_c.
\label{S4}
\end{equation}

\subsection{Series Expansion for Mean Cluster Number $S$}

As indicated in the Introduction, our ensuing analysis
requires partial knowledge of the series  expansion of the mean cluster number $S(p;d)$
for any dimension $d$ in powers of $p$:
\begin{equation}
S(p;d)=1+\sum_{m=1}  S_{m+1}(d) \, p^m.
\label{S5}
\end{equation}
The $d$-dependent coefficients $S_{k}(d)$, which account for ($k+1$)-mer cluster
configurations ($k=1,2,3,\cdots$), can be obtained in a number of different ways.
A common way is to first obtain explicit formulas for the cluster size
distribution $n_k$ and then employ (\ref{S4}) to get the $p$ expansion of $S$  and thus the coefficients
$S_{m+1}$ of series (\ref{S5}) \cite{Do61,Es72,Sy76,Lor00}.
The cluster size distribution can generally be represented by
the following relation:
\begin{equation}
n_k = \sum_{k=1} g_{km}\; p^k(1-p)^m,
\label{sum2}
\end{equation}
where $g_{km}$ is the number of cluster configurations (lattice
animals) with size $k$ and perimeter $m$ associated with that
cluster size \cite{St92}. The basic calculation reduces to the
determination of $g_{km}$. In Appendix A, we provide an algorithm
that enables one to obtain the explicit analytical expressions for
the $n_1$, $n_2$, $n_3$ and $n_4$ in arbitrary dimension for both
site and bond percolation for various $d$-dimensional lattices.

Another procedure that has been employed to ascertain the series (\ref{S5})
is to make use of the Mayer-type expansion of the pair connectedness
function $P({\bf r})$ in terms of the connectivity function $f({\bf r})$
defined by (\ref{mayer}) \cite{Co77}. In order to make contact with the
techniques used in Ref. \onlinecite{To12a} for continuum percolation,
it is useful here to map those results for the Mayer-type expansion of $P({\bf r})$
into the appropriate results for lattice percolation. For this purpose,
this mapping, which amounts to replacing integrals given in Ref. \onlinecite{To12a} with appropriate sums,
yields the following expansion of $P({\bf r})$ to first order in $p$ for lattice percolation:
\begin{equation}
P({\bf r}_{12}) = f({\bf r}_{12}) + p \; [1- f({\bf r}_{12})]\sum_{j} f({\bf r}_{1j})f({\bf r}_{2j}) + {\cal O}(p^2).
\label{P2}
\end{equation}
Substitution of (\ref{P2}) into (\ref{S1}) yields, after comparison to (\ref{S5}),
the dimer coefficient as
\begin{equation}
S_2(d) = \sum_{j=1} f({\bf r}_{1j}) =z_\alpha,
\label{s2}
\end{equation}
where $\alpha = s$ or $b$ for site or bond percolation,
respectively, and is related to the coordination number
$z_{\Lambda}$ of the lattice $\Lambda$ via either (\ref{zs}) or
(\ref{zb}). Similarly, the trimer coefficient are given by
\begin{eqnarray}
S_3(d)&=& \sum_{k} \sum_{j} [1-f({\bf r}_{1k})] f({\bf r}_{1j}),
f({\bf r}_{kj}) \label{s3}
\end{eqnarray}
where the indices $j$ and $k$ run through all sites (bonds). The
expressions (\ref{s2}) and (\ref{s3}) for the dimer and trimer
coefficients are the lattice analogs of Eqs. (24) and (25) given
in Ref. \onlinecite{To12a} for continuum percolation. In Appendix
B, we illustrate how to apply Eq. (\ref{s3}) by explicitly
computing $S_3$ for site percolation on the triangular lattice in
$\mathbb{R}^2$ (i.e., $A_2^*$).



\section{Lower  Bounds on the Percolation Threshold}
\label{Lower}

It has recently been shown that the  $[0,1]$, $[1,1]$ and $[2,1]$ Pad{\'e } approximant
of the mean cluster number $S$, a function of the particle number density, for prototypical $d$-dimensional continuum
percolation models provide lower bounds on the corresponding thresholds \cite{To12a}.
Specifically, these results apply to overlapping (Poisson distributed) hyperspheres
as well as hyperparticles of nonspherical shapes with some specified orientational
distribution function. Using the same techniques as was employed in Ref. \onlinecite{To12a}, we obtain analogous lower bounds on $p_c$ for site and bond percolation for general $d$-dimensional lattices in $\mathbb{R}^d$.

Let us denote the $[n,1]$ Pad{\'e } approximant of the
series expansion (\ref{S5}) of the mean cluster number $S$ by $S_{[n,1]}$.
This rational function for any $d$ is given explicitly by
\begin{equation}
S \approx S_{[n,1]} =\frac{\displaystyle 1+\sum_{m=1}^n \left[S_{m+1} - S_m \frac{S_{n+2}}{S_{n+1}} \right]p^{m}}
{\displaystyle1-\frac{S_{n+2}}{S_{n+1}} p}, \qquad \mbox{for} \quad 0 \le p \le p_0^{(n)},
\label{n-1}
\end{equation}
where $p_0^{(n)}$ is the pole of the $[n,1]$ approximant, which is given by
\begin{equation}
p_0^{(n)} = \frac{S_{n+1}}{S_{n+2}}, \qquad \mbox{for} \quad n \ge 0,
\label{pole-pade}
\end{equation}
and $S_0 \equiv 1$. Here we use the convention that the sum in (\ref{n-1}) is zero in the single instance $n=0$.
The claim that we make is that the pole $p_0^{(n)}$ for $n=0,1$ and 2 bounds the threshold $p_c$
for general $d$-dimensional lattice percolation (site or bond) from
below for any $d$, i.e.,
\begin{equation}
p_c \ge  p_0^{(n)} = \frac{S_{n+1}}{S_{n+2}},  \qquad \mbox{for} \quad n=0,1,2.
\label{lower}
\end{equation}
For the $[n,1]$  Pad{\' e} bounds to become progressively better as
$n$ increases from 0 to 1 and then to 2, it is clear that the following conditions must be
obeyed:
\begin{equation}
S_2^2 \ge S_3, \qquad S_3^2 \ge S_2 S_4. \label{conditions}
\end{equation}

\subsection{Proof in the One-Dimensional Case}
\label{1D-bounds}

For the one-dimensional integer lattice $\mathbb{Z}$, it is trivial to show that
all $[n,1]$ Pad{\' e} approximants of $S$ ($n=0,1,2,3,\ldots$)
provide lower bounds on the percolation threshold. To see this, note
the mean cluster number $S$ in this one-dimensional case is given exactly by
\begin{equation}
S=\frac{1+p}{1-p},
\label{pole-1}
\end{equation}
and hence the percolation threshold is trivially $p_c=1$.
Expanding this relation in powers of $p$ and comparing  to (\ref{S5}) yields
\begin{equation}
S_m= 2 \;, \quad \mbox{for}\quad m \ge 2.
\end{equation}
We see from (\ref{pole-pade}) that
\begin{equation}
p_0^{(n)}= \Bigg\{{1/2 \quad \mbox{for $n=0$},\atop{1, \quad \mbox{for $n \ge 1$.}}}
\end{equation}
and hence these poles always bound from below or equal the actual threshold $p_c=1$.

{\it Remark}:
For sufficiently small $d \ge 2$, all $[n,1]$ Pad{\' e} approximants
of $S$ ($n=0,1,2,3,\ldots$) cannot be nontrivial positive upper bounds on $S$.
For example, it is known that for $d=2$, $S_m$ can be negative
for some sufficiently large $m$ \cite{Sy73}.

\subsection{$[0,1]$ Pad{\' e} Bounds}
\label{zero-one}

We will begin by proving that the $[0,1]$ Pad{\'e } approximant of the mean cluster number,
\begin{equation}
S \approx S_{[0,1]} =\frac{1}{\displaystyle1- S_2 \,p}=\frac{1}{\displaystyle 1-\frac{p}{z_\alpha}}, \qquad \mbox{for} 
\quad 0 \le p \le {z_{\alpha}^{-1}},
\label{0-1}
\end{equation}
provides the following rigorous lower bound on the percolation threshold $p_c$ for all $d$:
\begin{equation}
p_c \ge p_0^{(0)}=\frac{1}{z_{\alpha}},
\label{eta-0-1}
\end{equation}
where we have used the identity $S_2=z_{\alpha}$ [cf. (\ref{s2})] and 
$z_\alpha$ is given by $z_\Lambda$ [cf. (\ref{zs})] and $2(z_\Lambda-1)$ [cf. (\ref{zb})]
for site and bond percolation, respectively. It follows that
in the high-$d$ limit, the pole $p_0^{(0)}$ for site percolation is twice
that for  bond percolation on some $d$-dimensional lattice, as reflected
in the asymptotic expansions given in Sec. \ref{Exact} for specific lattices.

Here we follow the analogous proof given for continuum percolation given in
Ref. \onlinecite{To12a} using the aforementioned mapping between the continuum and
lattice problem. In particular, bounds (100) and (101) for the pair connectedness
function $P({\bf r})$ given in that paper become for lattice percolation
\begin{eqnarray}
P({\bf r}_{12})&\ge & f({\bf r}_{12}), \label{1} \\
P({\bf r}_{12})&\le & f({\bf r}_{12}) + p \; [1- f({\bf r}_{12})]\sum_{j} f({\bf r}_{1j})P({\bf r}_{2j}). \label{2}
\end{eqnarray}
Note the similarity of the lower bound (\ref{2}) to the low-$p$ expansion (\ref{P2});
except here $P$ replaces $f$ in the sum and inequality (\ref{2})
is valid for arbitrary $p$. Note that since $1-f({\bf r}) \le 1$, we also have from (\ref{2}), the weaker
upper bound
\begin{eqnarray}
P({\bf r}_{12})&\le & f({\bf r}_{12}) + p \; \sum_{j} f({\bf r}_{1j})P({\bf r}_{2j}). \label{3}
\end{eqnarray}
Summing  inequality (\ref{3}) over site (bond) 2 and using the
definition (\ref{S2}) for the mean cluster number $S$ yields the
following upper bound on the latter:
\begin{equation}
S \le \frac{1}{1-S_2 \,p}.
\label{bound1}
\end{equation}
Now since this lower bound has a pole at $p=S_2^{-1}=z_{\alpha}^{-1}$, it
immediately implies the new rigorous lower bound on the percolation
threshold (\ref{eta-0-1}) for any $d$. It is important to note
that this lower bound is valid for {\it any} $d$-dimensional lattice $\Lambda$.

Note that a stronger rigorous upper bound on $P({\bf r})$ can be obtained by using the lower bound (\ref{1})
in the inequality (\ref{2}), namely,
\begin{eqnarray}
P({\bf r}_{12})&\le & f({\bf r}_{12}) + p \; [1- f({\bf r}_{12})]\sum_{j} f({\bf r}_{1j})f({\bf r}_{2j}). \label{4}
\end{eqnarray}
Summing inequality (\ref{4}) over site 2 and use of
(\ref{S2}) and (\ref{s3}) gives the following upper bound:
\begin{equation}
S \le \frac{1+(S_2^2-S_3)p^2}{1-S_2\, p}.
\label{S-1-1}
\end{equation}
Although this lower bound on $S$ is sharper than (\ref{bound1}), it has the
same pole and therefore does not provide a tighter upper bound
on the percolation threshold than (\ref{eta-0-1}).

\subsection{$[1,1]$ and $[2,1]$ Pad{\' e} Bounds}

The $[1,1]$ Pad{\'e } approximant of  $S$, given by (\ref{n-1}) with $n=1$, is more explicitly
given by
\begin{equation}
S \le S_{[1,1]} =\frac{\displaystyle 1+\left[z_{\alpha}-\frac{S_{3}}{z_{\alpha}}\right]p}{\displaystyle 1-\frac{S_3}{z_{\alpha}} p}, \qquad \mbox{for} \quad 0 \le p \le p_0^{(1)},
\label{1-1}
\end{equation}
provides the following putative lower bound on the  threshold $p_c$ in all Euclidean dimensions:
\begin{equation}
p_c \ge p_0^{(1)}=\frac{\displaystyle z_{\alpha}}{\displaystyle S_3},
\label{eta-1-1}
\end{equation}
where $p_0^{(1)}$ is the pole defined by (\ref{pole-pade}) and we have made use of the
identity $S_2=z_{\alpha}$.

Aizenman and Newman \cite{Ai84} used completely different methods
to prove, for the special case of {\it bond} percolation on the
hypercubic lattice $\mathbb{Z}^d$, the following upper bound on $S$:
\begin{equation}
S \le \frac{1}{\displaystyle 1 - 2d\,p}
\label{A1}
\end{equation}
and hence
\begin{equation}
p_c \ge \frac{1}{2d}.
\label{A2}
\end{equation}
It is instructive to compare these bounds (that apply only for $\mathbb{Z}^d$)
to the [1,1] estimates. Using the fact that $S_2(d)= z_b=2(2d-1)$ and $S_3(d)=2(2d-1)^2$ for 
bond percolation on the hypercubic lattice  (see results of Sec. \ref{expansions}), the
[1,1] estimates (\ref{1-1}) and (\ref{eta-1-1}) reduce to
\begin{equation}
S \le \frac{1}{\displaystyle 1 - (2d-1)p},
\label{C1}
\end{equation}
\begin{equation}
p_c \ge \frac{1}{2d-1}.
\label{C2}
\end{equation}
It is seen that the [1,1] estimates (\ref{C1}) and (\ref{C2}) for
the special case of bond percolation on $\mathbb{Z}^d$ provide 
sharper bounds than (\ref{A1}) and (\ref{A2}) in any finite dimension,
and tend to the same asymptotic bound in the limit $d \rightarrow \infty$.

Similarly, the $[2,1]$ Pad{\'e } approximant of the mean cluster number $S$,  given by (\ref{n-1}) with $n=2$, is more explicitly
given by
\begin{equation}
S \le S_{[1,1]} =\frac{\displaystyle 1+\left[z_{\alpha}-\frac{S_4}{S_3}\right] p
+\left[S_3-\frac{z_{\alpha} S_4}{S_3}\right] p^2}{\displaystyle 1-\frac{S_4}{S_3} p
}, \qquad \mbox{for} \quad 0 \le p \le p_0^{(2)},
\label{2-1}
\end{equation}
provides the following putative lower bound on the percolation threshold $p_c$ in all $d$:
\begin{equation}
p_c \ge p_0^{(2)} = \frac{S_3}{S_4},
\label{eta-2-1}
\end{equation}
where $p_0^{(2)}$ is the pole defined by (\ref{pole-pade}).
Since the expansion of upper bound (\ref{2-1}) in powers
of $p$ is exact through order $p^3$, we deduce, after comparison
to the  exact expansion (\ref{S5}), the following upper
bound on the fifth-order coefficient $S_5(d)$ for any
$d$-dimensional lattice $\Lambda$:
\begin{equation}
S_5(d) \le \frac{S_4^2(d)}{S_3(d)}.
\end{equation}

With considerably extra effort, one can rigorously prove that
(\ref{eta-1-1}) and (\ref{eta-2-1}) are indeed lower bounds on the
threshold $p_c$. However, this is beyond the scope of the present
paper, and will be reserved for a future work. Nonetheless, it is
noteworthy that high-dimensional asymptotic expansions of
(\ref{eta-1-1}) and (\ref{eta-2-1}) for both site and bond
percolation on the hypercubic lattice $\mathbb{Z}^d$ provide
lower bounds on the corresponding exact asymptotic expansions, as
explicitly shown in Sec. \ref{Exact-Bravais}. Moreover, in Sec. \ref{eval}, we will see that
available high-precision numerical estimates of $p_c$ for 
different lattices across dimensions support
the proposition that (\ref{eta-1-1}) and (\ref{eta-2-1})
are rigorous lower bounds on $p_c$.

\subsection{$[n,1]$ Pad{\' e} Approximant }

We expect that higher-order $[n,1]$ Pad{\' e} approximants ($n \ge 3$) of $S$ also provide
lower bounds on $p_c$ for $d\ge 2$ for $n \ge 3$
and relatively low $d$ provided that certain conditions are met.
One such necessary conditions is that successive coefficients $S_{n+1}$ and $S_{n+2}$ remain positive. 
For example, we have directly verified that both $S_{[3,1]}$ and $S_{[4,1]}$ yield lower bounds on $p_c$ for
$d=2$ and $d=3$ for a variety of site and bond problems on a
variety of lattices \cite{Do61,Es72,Sy76,Lor00}. However, as noted earlier, because we expect
$S_n$ to become negative at some
sufficiently large value of $n$ for $d=2$ and $d=3$, $S_{[n,1]}$ cannot always yield lower bounds
on $p_c$ for relatively low dimensions such that $d \ge 2$. In the limit
$d\rightarrow \infty$, we have shown that the $S_n$ are all positive 
and hence it is possible that in sufficiently high but finite $d$, $S_{[n,1]}$
gives lower bounds on $p_c$ for any $n$. The reader is referred to
a related discussion in Sec. \ref{discuss}.

\section{Series Expansions of $S$ for Various $d$-Dimensional Lattices}
\label{expansions}

\subsection{Definitions of the $d$-dimensional Lattices of Interest}

In this work, we consider the $d$-dimensional generalizations of the
simple-cubic lattice or simply hypercubic lattice $\mathbb{Z}^d$
as well as $d$-dimensional generalizations
of the face-centered-cubic, body-centered-cubic,
diamond and kagom{\'e } lattices  for $d\ge2$. While the first three are Bravais
lattices, the last two are non-Bravais lattices, as defined
more precisely below. It is noteworthy that  generalizations of these lattices
are {\it not} unique in higher dimensions.

\subsubsection{$d$-Dimensional Bravais Lattices}

The {\it hypercubic} $\mathbb{Z}^d$ is defined by
\begin{equation}
\mathbb{Z}^d=\{(x_1,\ldots,x_d): x_i \in {\mathbb{ Z}}\} \quad \mbox{for}\; d\ge 1
\end{equation}
where $\mathbb{Z}$ is the set of integers ($\ldots -3,-2,-1,0,1,2,3\ldots$)
and $x_1,\ldots,x_d$ denote the components of a lattice vector.
The coordination number of $\mathbb{Z}^d$ is $z_{\mathbb{Z}^d}=2d$.

A $d$-dimensional generalization of the face-centered-cubic lattice
is  the {\it checkerboard} $D_d$ lattice defined by
\begin{equation}
D_d=\{(x_1,\ldots,x_d)\in \mathbb{Z}^d: x_1+ \cdots +x_d ~~\mbox{even}\} \quad \mbox{for}\; d\ge 3.
\end{equation}
Its coordination number  is $z_{D_d}=2d(d-1)$.
Note that $D_2$ is simply the square lattice in
$\mathbb{R}^2$. The checkerboard lattice $D_d$ gives the densest sphere packing
for $d=3$ and the densest known sphere packings for $d = 4$ and 5, but not for higher dimensions \cite{Co93, Co03,To10d}.
It also provides the optimal {\it kissing-number} configurations for $d=3-5$,
but not for $d \ge 6$ \cite{Cohn11}.

In order to define the generalization of the body-centered-cubic
lattice that we will consider in this paper, we must first introduce
another generalization of the face-centered-cubic lattice, namely, the
{\it root} lattice $A_d$, which is a subset of points in $\mathbb{Z}^{d+1}$, i.e.,
\begin{equation}
A_d=\{(x_0,x_1,\ldots,x_d)\in \mathbb{Z}^{d+1}: x_0+ x_1+ \cdots +x_d =0\} \quad \mbox{for}\; d\ge 1.
\end{equation}
The coordination number of $A_d$ is $z_{A_d}=d(d+1)$. Note that $D_3=A_3$, but $D_d$ and $A_d$ are not the same lattices for $ d\ge 4$.
It is important to stress that the fundamental cell for the $A_d$ lattice
is a regular {\it rhombotope}, the $d$-dimensional generalization of the
two-dimensional rhombus or three-dimensional rhombohedron.

The $d$-dimensional lattices $\mathbb{Z}^d_*$, $D_d^*$ and $A_d^*$
are the corresponding dual lattices of $\mathbb{Z}^d$, $D_d$ and
$A_d$. While both $D_3^*$ and $A_3^*$ are the body-centered cubic
lattice, they are not the same lattices for $d \ge 4$. Indeed,
$D_d^*$ has an unusual coordination structure for $d \ge 4$ in
that the coordination number does  not increase monotonically with
$d$. By contrast, the coordination number of $A_d^*$ is $z_{A^*_d}=2(d+1)$.
For this reason, we choose to consider the $A_d^*$ lattice as a
$d$-dimensional generalization of the body-centered-cubic lattice.
The lattice vectors ${\bf e}_i$ of $A_d^*$ can be obtained from
the associated Gram matrix ${\bf G} = \{G_{ij}\} = <{\bf e}_i,
{\bf e}_j>$, where $<, >$ denotes the inner product of two vectors
in $\mathbb{R}^d$. Following Conway and Sloane, we set $G_{ii} =
d$ and $G_{ij} = -1$ ($i \neq j$). We note that $A_2^*\equiv A_2$ is
the triangular lattice in $\mathbb{R}^2$. (We say that two
lattices are {\it equivalent} or {\it similar} if one becomes
identical to the other by possibly rotation, reflection, and
change of scale, for which we use the symbol $\equiv$.) The
lattice $A_d^*$ provides the best known {\it covering} of
$\mathbb{R}^d$ in dimensions 1-5 and 10-18 \cite{Co93, To10d}.
We note that while $A_3^*$ apparently minimizes
large-scale density fluctuations (among all point configurations in $\mathbb{R}^d$),
this is not true for the corresponding problem for $d=4$,
where $D_4^*$ is the best known solution \cite{To10d}.

\subsubsection{$d$-Dimensional Non-Bravais Lattices}
\label{non}

The generalizations of the diamond and kagom{\'e} lattices
considered here were introduced in Ref. \onlinecite{Za09}. Specifically, since
the fundamental cell for the $A_d$ lattice is a regular
rhombotope, the points $\{{\bf 0}\}\cup \{{\bf a}_j\}$ ($j=1,
\ldots, d$), where ${\bf a}_j$ denotes a lattice vector of $A_d$,
are situated at the vertices of a regular $d$-dimensional simplex.
The $d$-dimensional diamond lattice $\mbox{Dia}_d$ can be obtained by including
in the fundamental cell the centroid of this simplex, i.e.,
\begin{equation}
{\boldsymbol \nu} = \frac{1}{d+1}\sum_{j=1}^d {\bf a}_j,
\end{equation}
which leads to a lattice with two basis points per fundamental
cell. By construction, the number of nearest neighbors to each
point in $\mbox{Dia}_d$ is $z_{\scriptsize \mbox{Dia}_d}=d + 1$,
corresponding to one neighbor for each vertex of a regular
$d$-simplex ($d$-dimensional generalization of the tetrahedron). Note that $\mbox{Dia}_2$ is the
usual honeycomb lattice, in which each point is at the vertex of a
regular hexagon.

Similar to the construction of the $d$-dimensional diamond
lattice, the $d$-dimensional kagom{\'e} lattice $\mbox{Kag}_d$ can be obtained by
placing lattice points at the midpoints of each nearest-neighbor
bond in the $A_d$ lattice \cite{Za09}. With respect to the underlying $A_d$
lattice, these lattice points are located at
\begin{equation}
\begin{array}{c}
{\bf x}_0 = {\boldsymbol \nu}/2 \\
{\bf x}_j = {\boldsymbol \nu}+{\boldsymbol p}_j/2
\end{array}
\end{equation}
where ${\boldsymbol p}_j = {\bf a}_j -{\boldsymbol \nu}$. By
translating the fundamental cell such that the origin is at ${\bf
x}_0$, we can also represent $\mbox{Kag}_d$ as $A_d\oplus\{{\bf v}_j\}$, where ${\bf v}_j = {\bf a}_j
/2$ $(j = 1, \ldots d)$. $\mbox{Kag}_d$
has $d+1$ basis points per fundamental cell, growing linearly with
dimension. Each lattice site is at the vertex of a regular simplex
obtained by connecting all nearest neighbors in the lattice,
implying that each point possesses $2d$ nearest neighbors in
$\mathbb{R}^d$, i..e, $z_{\scriptsize\mbox{Kag}_d}=2d$. We note that our $d$-dimensional kagom{\'e}
lattice is equivalent to the construction discussed in Ref. \onlinecite{Va98a}.

\subsection{Analytical Formulas for the Coefficients $S_2(d)$,  $S_3(d)$ and $S_4(d)$}
\label{Sk}


Here we provide [using the cluster-size distribution function $n_k$ expressions 
given in Appendix A and Eq. (\ref{S4})] explicit analytical formulas for the
$d$-dimensional coefficients $S_2(d)$,  $S_3(d)$ and $S_4(d)$ associated with
the series expansion of $S$ in powers of $p$ [cf. (\ref{S5})] for general dimension
$d$ in the cases of the $\mathbb{Z}^d$, $D_d$, $A_d^*$, $\mbox{Dia}_d$ and
$\mbox{Kag}_d$ lattices for both site and bond percolation. In seven
out of these ten problems, such $d$-dimensional expansions
have heretofore not been given. These coefficients together with the
general lower bounds given in Sec. ~\ref{Lower} give
corresponding explicit lower bounds on $p_c$ for these
ten percolation problems.

For the hypercubic lattice $\mathbb{Z}^d$, the series expansion of
$S$ in powers of $p$ for site and percolation,
through third-order in $p$, are given respectively  by
\begin{equation}
S = 1 + 2d p+ 2d(2d-1) p^2 + 2d(4d^2-7d+4) p^3 + {\cal O}(p^4),
\label{CS}
\end{equation}
\begin{equation}
S = 1 + 2(2d-1) p+ 2d(2d-1)^2 p^2 + 2(8d^3-12d^2+3d+2) p^3 +
{\cal O}(p^4).
\label{CB}
\end{equation}
The results (\ref{CS}) and (\ref{CB}) agree with earlier ones
reported in Refs. \onlinecite{Ga76} and \onlinecite{Ga78}, respectively.

For the $d$-dimensional checkerboard lattice $D_d$ (the
generalization of the fcc lattice), the series expansion of $S$
for site and bond percolation are given respectively  by
\begin{equation}
S = 1 + 2d(d-1) p+ 2d(d-1)(2d^2-6d+7) p^2 + 2d(d-1)(4d^4 - 24d^3 +
57d^2 - 53d + 12) p^3 + {\cal O}(p^4),
\end{equation}
\begin{equation}
S = 1 + 2(2d^2-2d-1) p+ 2(4d^4-8d^3+9) p^2 +
2(8d^6-24d^5+12d^4-8d^3+27d^2+131d-218) p^3 + {\cal O}(p^4).
\end{equation}


For $\mathbb{A}^*_d$ (our $d$-dimensional generalization of the
bcc lattice), the series expansion of $S$ for site and bond percolation 
are given respectively  by
\begin{equation}
S = \left \{ { 
\begin{array}{c}
1 + 6 p + 18 p^2 + 48 p^3+ {\cal O}(p^4), \quad d=2 \\
1 + 8 p + 56 p^2 + 248 p^3+ {\cal O}(p^4), \quad d=3 \\
1 + 2(d+1) p+ 2(d+1)(2d+1)
p^2 + 2(d+1)(4d^2+d+1) + {\cal O}(p^4), \quad d\ge4
\end{array}
} \right.
\end{equation}
\begin{equation}
S = \left \{ { 
\begin{array}{c}
1 + 10 p + 46 p^2 + 186 p^3+ {\cal O}(p^4), \quad d=2 \\
1 + 14 p + 98 p^2 + 650 p^3+ {\cal O}(p^4), \quad d=3 \\
1 + 2(2d+1) p+ 2(2d+1)^2 p^2 + 2(8d^3+12d^2+3d+1) p^3 +
{\cal O}(p^4), \quad d\ge4
\end{array}
} \right.
\end{equation}


For the $d$-dimensional diamond lattice $\mbox{Dia}_d$, the series
expansion of $S$ for site and bond percolation are given respectively  by
\begin{equation}
S = 1 + (d+1) p+ d(d+1) p^2 + d^2(d+1) p^3 + {\cal O}(p^4),
\end{equation}
\begin{equation}
S = 1 + 2d p+ 2d^2 p^2 + 2d^3 p^3 + {\cal O}(p^4).
\end{equation}

For the $d$-dimensional kagom{\'e} lattice $\mbox{Kag}_d$, the
series expansion of $S$ for site and bond percolation are given respectively  by
\begin{equation}
S = 1 + 2d p+ 2d^2 p^2 + 2d^3 p^3 + {\cal O}(p^4),
\end{equation}
\begin{equation}
S = 1 + 2(2d-1) p+ 2(4d^2-5d+2) p^2 + (16d^3-39d^2+43d-18) p^3 +
{\cal O}(p^4).
\end{equation}
The expansion for site percolation agrees with the one first reported in Ref. \onlinecite{Va98a}.

\begin{table}[!htp]
\caption{The $d$-dependent coefficients $S_{k}(d)$ for site
percolation. For $A^*_d$, the expressions apply in dimensions four
and higher. For all of the other lattices, the expressions apply
in dimensions two and higher.}
\begin{tabular}{c|c|c|c }
\hline
Lattice& $S_2(d)$ & $S_3(d)$ & $S_4(d)$   \\
\hline
$\mathbb{Z}^d$ &   $2d$ &  $2d(2d-1)$ & $2d(4d^2-7d+4)$ \\
$\mathbb{D}_d$ &  ~$2d(d-1)$~ &  ~$2d(d-1)(2d^2-6d+7)$~ & ~$2d(d-1)(4d^4 - 24d^3 + 57d^2 - 53d + 12)~$ \\
$\mathbb{A}^*_d$ &   $2(d+1)$ &  $2(d+1)(2d+1)$ & $2(d+1)(4d^2+d+1)$ \\
$\mbox{Dia}_d$ &  $d+1$ &  $d(d+1)$ & $d^2(d+1)$ \\
$\mbox{Kag}_d$ &  $2d$ &  $2d^2$ & $2d^3$ \\
\hline
\end{tabular}
\label{site}
\end{table}

The $d$-dependent coefficients $S_{k}(d)$ are also summarized in
Tables~\ref{site} and \ref{bond} for site and bond percolation
respectively for various $d$-dimensional lattices. We note
that the coefficients $S_2(p)$, $S_3(p)$ and
$S_4(p)$ for all of the $d$-dimensional lattices summarized in
in these tables satisfy the conditions
(\ref{conditions}) and hence the [0,1], [1,1] and [2,1] lower
bounds on $p_c$ progressively improve as the order increases.
Since nearest neighbors sites in
$\mbox{Kag}_d$ correspond exactly to nearest neighbor bonds in
$\mbox{Dia}_d$, it is not surprising that the
coefficients $S_k(d)$ for {\it site} percolation on $\mbox{Kag}_d$
and those for {\it bond} percolation on $\mbox{Dia}_d$ are
identical, as shown here.


\begin{table}[!htp]
\caption{The $d$-dependent coefficients $S_{k}(d)$ for bond
percolation. For $A^*_d$, the expressions apply in dimensions four
and higher. For all of the other lattices, the expressions apply
in dimensions two and higher.}
\begin{tabular}{c|c|c|c }
\hline
Lattice& $S_2(d)$ & $S_3(d)$ & $S_4(d)$   \\
\hline
$\mathbb{Z}^d$ &   $2(2d-1)$ &  $2(2d-1)^2$ & $2(8d^3-12d^2+3d+2)$ \\
$\mathbb{D}_d$ &  ~$2(2d^2-2d-1)$~ &  ~$2(4d^4-8d^3+9)$~ & ~$2(8d^6-24d^5+12d^4-8d^3+27d^2+131d-218)$~ \\
$\mathbb{A}^*_d$ &   $2(2d+1)$ &  $2(2d+1)^2$ & $2(8d^3+12d^2+3d+1)$ \\
$\mbox{Dia}_d$ &  $2d$ &  $2d^2$ & $2d^3$ \\
$\mbox{Kag}_d$ &  $ 2(2d-1)$ &  $2(4d^2-5d+2)$ & $(16d^3-39d^2+43d-18)$ \\
\hline
\end{tabular}
\label{bond}
\end{table}

\subsection{Exact High-$d$ Asymptotics for the Percolation Threshold $p_c$ }
\label{Exact}

Here we obtain the high-dimensional asymptotic expansions of the
lower bounds on $p_c$ that were obtained from the [0,1], [1,1] and
[2,1] Pad{\'e} approximants of $S$ for the hypercubic lattice
$\mathbb{Z}^d$ as well as $d$-dimensional generalizations of the
face-centered-cubic ($D_d$), body-centered-cubic ($A_d^*$),
diamond ($\mbox{Dia}_d$), and kagom{\'e} ($\mbox{Kag}_d$)
lattices. While we show that eight of the ten asymptotic
expansions agree with the high-dimensional Bethe approximation (\ref{B2}),
corresponding results for site percolation on both  $\mbox{Dia}_d$ and $\mbox{Kag}_d$ do not.

\subsubsection{$d$-Dimensional Bravais Lattices $\mathbb{Z}^d$, $D_d$ and $A_d^*$
}
\label{Exact-Bravais}

In the case of site percolation on the hypercubic lattice
$\mathbb{Z}^d$, the high-dimensional asymptotic
expansions of the lower bounds (\ref{eta-0-1}),  (\ref{eta-1-1})
and  (\ref{eta-2-1}) on $p_c$ obtained from the [0,1],
[1,1] and [2,1] Pad{\'e} approximants of $S$ are respectively
given by:
\begin{eqnarray}
p_c &\ge& \frac{1}{2d} \label{CS1} \\
p_c  &\ge& \frac{1}{2d} +\frac{1}{4d^2}  +\frac{1}{8d^3} + {\cal O}\left(\frac{1}{d^4}\right) \label{CS2}\\
p_c  &\ge& \frac{1}{2d} +\frac{5}{8d^2}  +\frac{19}{32d^3} + {\cal O}\left(\frac{1}{d^4}\right) \label{CS3}
\end{eqnarray}
This is to be compared to exact asymptotic expansion obtained by Gaunt, Sykes and Ruskin \cite{Ga76} by to the same
order:
\begin{eqnarray}
p_c  &=& \frac{1}{2d} +\frac{5}{8d^2}  +\frac{31}{32d^3} + {\cal O}\left(\frac{1}{d^4}\right).
\end{eqnarray}
The tightest lower bound is exact up through order
$1/d^2$ and its third-order coefficient $19/32$ bounds the exact
third-order coefficient $31/32$ from below, as expected. It is
noteworthy that the leading-order term in the exact asymptotic
expansion is inversely proportional to the coordination number
$z_{\mathbb{Z}^d}=z_s=2d$. This is consistent with the high-dimensional
Bethe approximation (\ref{B2}). Moreover, the leading-order term
in the asymptotic expansion, obtained from the [0, 1] lower bound
(i.e., $p_c \ge 1/S_2$), always agrees with the Bethe
approximation since $S_2 = z_{\mathbb{Z}^d}$ [c.f. Eq.~(\ref{s2})].
In the instance of bond percolation on the $\mathbb{Z}^d$, the
asymptotic expansions of the  [0,1], [1,1] [2,1] Pad{\'e} lower
bounds respectively yield
\begin{eqnarray}
p_c  &\ge& \frac{1}{4d} +\frac{1}{8d^2}  +\frac{1}{16d^3} +\frac{1}{32d^4} + {\cal O}\left(\frac{1}{d^5}\right)  \label{CB1} \\
p_c  &\ge& \frac{1}{2d} +\frac{1}{4d^2}  +\frac{1}{8d^3} + \frac{1}{16d^4} +{\cal O}\left(\frac{1}{d^5}\right)  \label{CB2} \\
p_c  &\ge& \frac{1}{2d} +\frac{1}{4d^2}  +\frac{5}{16d^3}
+\frac{1}{4d^4} + {\cal O}\left(\frac{1}{d^5}\right) \label{CB3}
\end{eqnarray}
These results are be compared to the exact asymptotic expansion obtained by
Gaunt and Ruskin \cite{Ga78} to the same order:
\begin{eqnarray}
p_c  &=& \frac{1}{2d} +\frac{5}{4d^2}  +\frac{19}{16d^3} +\frac{1}{d^4} + {\cal O}\left(\frac{1}{d^5}\right).
\end{eqnarray}
Observe that the tightest lower bound in the case of bond
percolation is exact up through order $1/d$ (in contrast to the
corresponding site percolation bound that is exact through order
$1/d^2$) and its second-order coefficient $1/4$ bounds the exact
second-order coefficient $5/4$ from below, as it should. As in the
case of site percolation on $\mathbb{Z}_d$, the leading order term
of the exact asymptotic expansion of $p_c$ for bond percolation on
this lattice agrees with the  Bethe approximation (\ref{B2})
[i.e., $p_c \sim 1/z_{\mathbb{Z}^d}=1/(2d)$].


In the case of site percolation on the checkerboard lattice
$\mathbb{D}_d$, the asymptotic expansions of the
obtained from the [0,1], [1,1] and [2,1] Pad{\'e} lower bounds on $p_c$
 respectively yield
\begin{eqnarray}
p_c  &\ge& \frac{1}{2d^2} +\frac{1}{2d^3}  +\frac{1}{2d^4}  + {\cal O}\left(\frac{1}{d^5}\right),  \label{FS1} \\
p_c  &\ge& \frac{1}{2d^2} +\frac{3}{2d^3}  +\frac{11}{4d^4}  +{\cal O}\left(\frac{1}{d^5}\right), \label{FS2} \\
p_c  &\ge& \frac{1}{2d^2} +\frac{3}{2d^3}  +\frac{29}{8d^4}  +{\cal O}\left(\frac{1}{d^5}\right). \label{FS3}
\end{eqnarray}
These results lead to the conclusion that the asymptotic
expansion of the tightest lower bound  is exact at least
through order $1/d^3$ and hence
\begin{equation}
p_c  = \frac{1}{2d^2} +\frac{3}{2d^3}  +{\cal O}\left(\frac{1}{d^4}\right)
\end{equation}
For bond percolation on $\mathbb{D}_d$, the asymptotic expansions of the lower bounds
yield
\begin{eqnarray}
p_c  &\ge& \frac{1}{4d^2} +\frac{1}{4d^3}  +\frac{3}{8d^4} +\frac{1}{2d^5} + {\cal O}\left(\frac{1}{d^6}\right),  \label{FB1} \\
p_c  &\ge& \frac{1}{2d^2} +\frac{1}{2d^3}  +\frac{3}{4d^4} + \frac{3}{2d^5} +{\cal O}\left(\frac{1}{d^6}\right),  \label{FB2} \\
p_c  &\ge& \frac{1}{2d^2} +\frac{1}{2d^3}  +\frac{3}{4d^4}
+\frac{2}{d^5} + {\cal O}\left(\frac{1}{d^6}\right).\label{FB3}
\end{eqnarray}
Thus, we see that these results lead to the conclusion that the asymptotic
expansion of the tightest lower bound  is exact at least
through order $1/d^4$, implying
\begin{equation}
p_c  = \frac{1}{2d^2} +\frac{1}{2d^3}  +\frac{3}{4d^4}
 + {\cal O}\left(\frac{1}{d^5}\right) .
\end{equation}
Note that the exact leading order terms of the asymptotic
expansions of $p_c$ for both site and bond percolation on $D_d$
agree with the  Bethe approximation (\ref{B2}) [i.e., $p_c
\sim 1/z_{D_d}=1/(2d^2)$].

In the case of site percolation on ${A}_d^*$, the asymptotic
expansions of the lower bounds obtained from the [0,1], [1,1] and
[2,1] Pad{\'e} approximants of $S$ respectively yield
\begin{eqnarray}
p_c  &\ge& \frac{1}{2d} -\frac{1}{2d^2}   + {\cal O}\left(\frac{1}{d^3}\right)  \label{AS1} \\
p_c  &\ge& \frac{1}{2d} -\frac{3}{4d^2}   +{\cal O}\left(\frac{1}{d^3}\right)  \label{AS2} \\
p_c  &\ge& \frac{1}{2d} +\frac{1}{8d^2}   +{\cal O}\left(\frac{1}{d^3}\right) \label{AS3}
\end{eqnarray}
These results lead to the conclusion that the asymptotic
expansion of the tightest lower bound  is exact at least
through order $1/d$ or, more precisely,
\begin{equation}
p_c  = \frac{1}{2d}  +{\cal O}\left(\frac{1}{d^2}\right)
\end{equation}
For bond percolation on ${A}_d^*$, the asymptotic expansions of the lower bounds
yield
\begin{eqnarray}
p_c  &\ge& \frac{1}{4d} -\frac{1}{8d^2}  +\frac{1}{16d^3} + {\cal O}\left(\frac{1}{d^4}\right)  \label{AB1} \\
p_c  &\ge& \frac{1}{2d} -\frac{1}{4d^2}  +\frac{1}{8d^3}  +{\cal O}\left(\frac{1}{d^4}\right)  \label{AB2} \\
p_c  &\ge& \frac{1}{2d} -\frac{1}{4d^2}  +\frac{5}{16d^3} + {\cal O}\left(\frac{1}{d^4}\right) \label{AB3}
\end{eqnarray}
Thus, we see that these results lead to the conclusion that the asymptotic
expansion of the tightest lower bound  is exact at least
through order $1/d^2$, and hence
\begin{equation}
p_c  =\frac{1}{2d} -\frac{1}{4d^2} + {\cal O}\left(\frac{1}{d^3}\right).
\end{equation}
As in all of the previous cases, we see that the exact leading
order terms of the asymptotic expansions of $p_c$ for both site
and bond percolation on $A_d^*$ agree with the high-$d$ Bethe
approximation (\ref{B2}) [i.e., $p_c \sim 1/z_{A^*_d}=1/(2d)$].

\subsubsection{$d$-Dimensional Non-Bravais Lattices $\mbox{Dia}_d$ and $\mbox{Kag}_d$
}

In the case of site percolation on the $d$-dimensional diamond
lattice $\mbox{Dia}_d$, all three lower bounds yield the same
asymptotic expansion, 
\begin{equation}
p_c  = \frac{1}{d} + \; \mbox{h.o.t},
\end{equation}
where $\mbox{h.o.t}$ indicates indeterminate higher-order terms.
For bond percolation on $\mbox{Dia}_d$, the asymptotic expansions
of the lower bounds obtained from the [0,1], [1,1] and [2,1]
Pad{\'e} approximants of $S$ respectively yield
\begin{eqnarray}
p_c  = \frac{1}{2d} + \; {\cal O}\left(\frac{1}{d^4}\right) \\
p_c  = \frac{1}{d} + \; {\cal O}\left(\frac{1}{d^4}\right) \\
p_c  = \frac{1}{d} + \; {\cal O}\left(\frac{1}{d^4}\right)
\end{eqnarray}
We know the order of the correction to the leading term since
this problem is identical to site percolation on
$\mbox{Kag}_d$ described below. The exact leading order terms for
both site and percolation on $\mbox{Dia}_d$ agree with the
Bethe approximation (\ref{B2}) [i.e., $p_c \sim
1/z_{\tiny\mbox{Dia}_d}=1/d$].


For site percolation on the $d$-dimensional kagom{\' e} lattice
$\mbox{Kag}_d$, the asymptotic expansions of the lower bounds
obtained from the [0,1], [1,1] and [2,1] Pad{\'e} approximants of
$S$ respectively yield
\begin{eqnarray}
p_c  = \frac{1}{2d} + \; {\cal O}\left(\frac{1}{d^4}\right) \\
p_c  = \frac{1}{d} + \; {\cal O}\left(\frac{1}{d^4}\right) \\
p_c  = \frac{1}{d} + \; {\cal O}\left(\frac{1}{d^4}\right)
\end{eqnarray}
We know the order of the correction to the leading term is ${\cal
O}(1/d^4)$, which we determined from the exact $p$-expansion of
$S$ through order $p^5$ obtained by van der Marck \cite{Va98a}. In
the case of bond percolation on $\mbox{Kag}_d$, the asymptotic
expansions of the three lower bounds yield
\begin{eqnarray}
p_c  &\ge& \frac{1}{4d} +\frac{1}{8d^2}  + {\cal O}\left(\frac{1}{d^3}\right)  \label{KB1} \\
p_c  &\ge& \frac{1}{2d} +\frac{3}{8d^2}  +{\cal O}\left(\frac{1}{d^3}\right)  \label{KB2} \\
p_c  &\ge& \frac{1}{2d} +\frac{19}{32d^2} + {\cal
O}\left(\frac{1}{d^3}\right) \label{KB3}
\end{eqnarray}
Note that these results lead to the conclusion that the asymptotic
expansion of the tightest lower bound  is exact at least
through order $1/d$, and hence
\begin{equation}
p_c  =\frac{1}{2d} + {\cal O}\left(\frac{1}{d^2}\right).
\end{equation}
While the asymptotic expansions of $p_c$ for bond
percolation on $\mbox{Kag}_d$ agree with the corresponding Bethe
approximation [i.e., $p_c \sim 1/z_{\tiny\mbox{Kag}_d}=1/(2d)$], this is not the 
case for site percolation [i.e., $p_c \sim 1/d \neq 1/z_{\tiny\mbox{Kag}_d}=1/(2d)$]. 
The latter observation was first made by van der Marck \cite{Va98a},
but no explanation for it was given. We will discuss this 
issue in Sec. VI.

\section{Evaluation of Bounds on $p_c$ and $S$, and Comparison to Simulation Results}
\label{eval}

Here, we explicitly evaluate the $[0,1]$, $[1,1]$
and $[2,1]$ lower bounds on $p_c$ [i.e., inequalities (\ref{eta-0-1}), (\ref{eta-1-1}) and  (\ref{eta-2-1})] for the
hypercubic lattice $\mathbb{Z}^d$ as well as $d$-dimensional
generalizations of the face-centered-cubic ($D_d$),
body-centered-cubic ($A^*_d$), kagom{\'e} ($\mbox{Kag}_d$), and
diamond lattices ($\mbox{Dia}_d$) up to dimension 13 using the results for
the corresponding coefficients $S_2(d)$, $S_3(d)$ and $S_4(d)$ listed
in Tables \ref{site} and \ref{bond}. We also employ
these results to ascertain the accuracy of previous numerical
simulations, especially in high dimensions. 

\begin{table}[!htp]
\caption{Comparison of numerical estimates of the site percolation
thresholds on the hypercubic lattice $\mathbb{Z}^d$ to
corresponding lower bounds on $p_c$. Simulation results for $d=2$,
$d=3$ and $d=4-13$ are taken from Refs. \onlinecite{Ne00},
\onlinecite{Sk09} and \onlinecite{Gr03}, respectively.}
\begin{tabular}{c|c|c|c|c }
\hline
~Dimension~& ~$p_c$~ & ~$p_c^L$ from Eq.~(\ref{eta-2-1})~ & ~$p_c^L$ from Eq.~(\ref{eta-1-1})~ & ~$p_c^L$ from Eq.~(\ref{eta-0-1})~  \\
\hline
1 & ~1.0000000000\ldots~ &  ~1.0000000000\ldots~ & ~1.0000000000\ldots~ & ~0.5000000000\ldots~ \\
2 & ~0.59274621~ & ~0.5000000000\ldots~ & ~0.3333333333\ldots~ & ~0.2500000000\ldots~\\
3 & ~0.3116004~ & ~0.2631578947\ldots~ & ~0.2000000000\ldots~ & ~0.1666666666\ldots~\\
4 & ~0.1968861~ & ~0.1750000000\ldots~ & ~0.1428571429\ldots~ & ~0.1250000000\ldots~\\
5 & ~0.1407966~ & ~0.1304347826\ldots~ & ~0.1111111111\ldots~ & ~0.1000000000\ldots~\\
6 & ~0.109017~ & ~0.1037735849\ldots~ & ~0.09090909090\ldots~ & ~0.08333333333\ldots~\\
7 & ~0.0889511~ & ~0.08609271523\ldots~ & ~0.07692307692\ldots~ & ~0.07142857143\ldots~\\
8 & ~0.0752101~ & ~0.07352941176\ldots~ & ~0.06666666666\ldots~ & ~0.06250000000\ldots~\\
9 & ~0.0652095~ & ~0.06415094340\ldots~ & ~0.05882352941\ldots~ & ~0.05555555555\ldots~\\
10 & ~0.0575930~ & ~0.05688622754\ldots~ & ~0.05263157895\ldots~ & ~0.05000000000\ldots~\\
11& ~0.05158971~ & ~0.05109489051\ldots~ & ~0.04761904762\ldots~ & ~0.04545454545\ldots~\\
12& ~0.04673099~ & ~0.04637096774\ldots~ & ~0.04347826087\ldots~ & ~0.04166666666\ldots~\\
13& ~0.04271508~ & ~0.04244482173\ldots~ & ~0.04000000000\ldots~ & ~0.03846153846\ldots~\\
\hline
\end{tabular}
\label{site-cubic}
\end{table}

\begin{figure}[htp]
\begin{center}
$\begin{array}{c}
\\\\
\includegraphics[height=6.0cm,keepaspectratio]{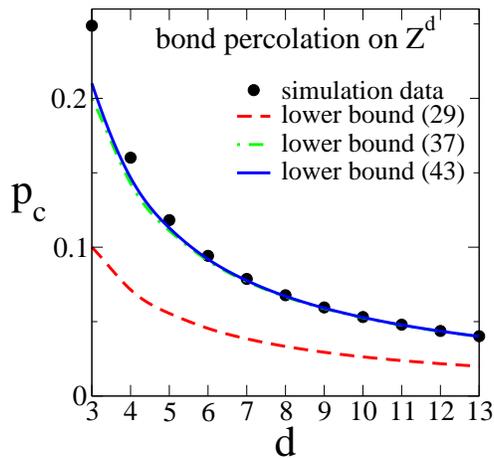}
\end{array}$
\end{center}
\caption{Percolation threshold $p_c$ versus dimension $d$ for bond
percolation on hypercubic lattice as obtained from the lower
bounds (\ref{eta-0-1}), (\ref{eta-1-1}), and (\ref{eta-2-1}) as
well as the simulation data.} \label{fig_hypercubic}
\end{figure}

In Tables \ref{site-cubic} and \ref{bond-cubic}, we compare the
lower bounds (\ref{eta-0-1}), (\ref{eta-1-1}) and (\ref{eta-2-1})
on the percolation threshold $p_c$ for site and bond percolation
on the hypercubic lattice $\mathbb{Z}^d$ up through dimension 13 to the corresponding
simulation data. It is can be clearly seen that the $[n, 1]$
Pad{\'e} bounds get progressively better as the order $n$
increases. Specifically, the $[2, 1]$ Pad{\'e} provides the
tightest lower bound on $p_c$, which becomes asymptotically exact
in the limit $d\rightarrow \infty$. The numerical values of $p_c$
for both site and bond percolation lie above the associated best
lower bound and approach the lower bound as $d$ increases,
indicating that these data are of high accuracy, as shown in  Figure
\ref{fig_hypercubic}. Assuming the level of accuracy claimed
in the simulations, our tightest lower bound
(\ref{eta-2-1}) is already accurate up to three significant figures
for $d \ge 10$.

\begin{table}[!htp]
\caption{Comparison of numerical estimates of the bond percolation
thresholds on the hypercubic lattice $\mathbb{Z}^d$ to
corresponding lower bounds on $p_c$. Simulation results for $d=3$
and $d=4-13$ are taken from Refs. \onlinecite{Lo98} and
\onlinecite{Gr03}, respectively.}
\begin{tabular}{c|c|c|c|c }
\hline
~Dimension~& ~$p_c$~ & ~$p_c^L$ from Eq.~(\ref{eta-2-1})~ & ~$p_c^L$ from Eq.~(\ref{eta-1-1})~ & ~$p_c^L$ from Eq.~(\ref{eta-0-1})~  \\
\hline
1 &  1 &  1 & 1 & 1/2 \\
2 & ~0.5000000000\ldots~ & ~0.3750000000\ldots~ & ~0.3333333333~ & ~0.1666666666\ldots~\\
3 & ~0.2488126~ & ~0.2100840336\ldots~ & ~0.2000000000\ldots~ & ~0.10000000000\ldots~\\
4 & ~0.1601314~ & ~0.1467065868\ldots~ & ~0.1428571429\ldots~ & ~0.07142857143\ldots~\\
5 & ~0.118172~ & ~0.1129707113\ldots~ & ~0.1111111111\ldots~ & ~0.05555555556\ldots~\\
6 & ~0.0942019~ & ~0.09194528875\ldots~ & ~0.09090909091\ldots~ & ~0.04545454545\ldots~\\
7 & ~0.0786752~ & ~0.07755851308\ldots~ & ~0.07692307692\ldots~ & ~0.03846153846\ldots~\\
8 & ~0.06770839~ & ~0.06708407871\ldots~ & ~0.06666666666\ldots~ & ~0.03333333333\ldots~\\
9 & ~0.05949601~ & ~0.05911229290\ldots~ & ~0.05882352941\ldots~ & ~0.02941176471\ldots~\\
10 & ~0.05309258~ & ~0.05283957845\ldots~ & ~0.05263157895\ldots~ & ~0.02631578947\ldots~\\
11& ~0.04794969~ & ~0.04777380565\ldots~ & ~0.04761904762\ldots~ & ~0.02380952381\ldots~\\
12& ~0.04372386~ & ~0.04359650569\ldots~ & ~0.04347826087\ldots~ & ~0.02173913043\ldots~\\
13& ~0.04018762~ & ~0.04009237283\ldots~ & ~0.04000000000\ldots~ & ~0.02000000000\ldots~\\
\hline
\end{tabular}
\label{bond-cubic}
\end{table}

\begin{table}[!htp]
\caption{Comparison of numerical estimates of the site percolation
thresholds on the checkerboard $D_d$ and $A_d^*$ lattices to the
corresponding best lower bounds on $p_c$. Simulation results for
$d=3$ and $d=4-6$ in the case of the  $D_d$ lattice are taken from
Refs. \onlinecite{Lor98} and \onlinecite{Va98b}, respectively.
Simulation results for $d=3$ in the case of the  $A_d^*$ or bcc
lattice is taken from Ref. \onlinecite{Lor98}.}
\begin{tabular}{c|c|c|c|c }
\hline
&  \multicolumn{2}{|c|}{$D_d$} & \multicolumn{2}{|c}{$A^*_d$} \\
~Dimension~& ~$p_c$~ & ~$p_c^L$ from Eq.~(\ref{eta-2-1}) & ~$p_c$~ & ~$p_c^L$ from Eq.~(\ref{eta-2-1}) \\
\hline
3 & ~0.1992365~ & ~0.1666666666\ldots~ & ~0.2459615~& ~0.2258064516\ldots~\\
4 & ~0.0842~ & ~0.07500000000\ldots~ & ~~ & ~0.1304347826\ldots~ \\
5 & ~0.0431~ & ~0.04017857143\ldots~& ~~ & ~0.1037735849\ldots~ \\
6 & ~0.0252~ & ~0.02462772050\ldots~& ~~ & ~0.08609271523\ldots~ \\
7 & ~~ & ~0.01655281135\ldots~& ~~             & ~0.07352941176\ldots~ \\
8 & ~~ & ~0.01186579378\ldots~& ~~             & ~0.06415094340\ldots~ \\
9 & ~~ & ~0.008914728682\ldots~& ~~             & ~0.05688622754\ldots~ \\
10 & ~~ & ~0.006939854594\ldots~ & ~~           & ~0.05109489051\ldots~\\
11& ~~ & ~0.005554543799\ldots~ & ~~            & ~0.04637096774\ldots~\\
12& ~~ & ~0.004545825179\ldots~ & ~~            & ~0.04244482173\ldots~\\
13& ~~ & ~0.003788738790\ldots~ & ~~            & ~0.03913043478\ldots~\\
\hline
\end{tabular}
\label{site-fcc}
\end{table}

We summarize in Table \ref{site-fcc} evaluations of the 
best lower bound (\ref{eta-2-1}) on the percolation threshold
$p_c$ for site percolation on the Bravais lattices $D_d$ and $A^*_d$ up through $d=13$
and compare them to corresponding simulation data when
available. Observe that (\ref{eta-2-1}) already provides a
tight bound on the numerical estimates of $p_c$ for $D_d$ in relatively low
dimensions (e.g., $d=5$ and 6). Our tightest lower bound (\ref{eta-2-1} ) estimates
for this lattice should provide sharp estimates of $p_c$
for $d \ge 6$ (where no numerical estimates
are currently available), which become progressively better as $d$ grows  
and, indeed, asymptotically exact in the high-$d$ limit.
In the case of $A_d^*$, only three-dimensional simulation
results are available for comparison.

The best lower bounds on $p_c$ for bond percolation on
the $D_d$ and $A^*_d$  lattices are compared
to available simulation data in Table \ref{bond-fcc}. Note that the
numerical estimates of $p_c$ for the $D_5$ lattice fall slightly below the
corresponding lower bound. This again illustrates the 
utility of tight bounds to assess the accuracy of numerical data
of $p_c$. We will see that the lower-bound 
estimate of $p_c$ for both site and bond percolation on $D_d$ converges to the corresponding
numerical estimates most rapidly among all of the $d$-dimensional
lattices that we have studied in this paper. The reasons for this
behavior are presented in Sec. \ref{discuss}.

\begin{table}[!htp]
\caption{Comparison of numerical estimates of the bond percolation
thresholds on the checkerboard ${D}_d$ and $A_d^*$ lattices to the
corresponding best lower bounds on $p_c$. Simulation results for
$d=3$ and $d=4-5$ in the case of the  $D_d$ lattice are taken from
Refs. \onlinecite{Lor98} and \onlinecite{Va98b}, respectively.
Simulation results for $d=3$ in the case of the  $A_d^*$ or bcc
lattice is taken from Ref. \onlinecite{Lor98}.}
\begin{tabular}{c|c|c|c|c }
\hline
&  \multicolumn{2}{|c|}{$D_d$} & \multicolumn{2}{|c}{$A^*_d$} \\
~Dimension~& ~$p_c$~ & ~$p_c^L$ from Eq.~(\ref{eta-2-1}) & ~$p_c$~ & ~$p_c^L$ from Eq.~(\ref{eta-2-1})  \\
\hline
3 & ~0.1201635~ & ~0.09965928450\ldots~ & ~0.1802875~& ~0.1467065868\ldots~\\
4 & ~0.0490~ & ~0.04534377720\ldots ~ & ~~ & ~0.1129707113\ldots~ \\
5 & ~0.026~ & ~0.02619245990\ldots~& ~~ & ~0.09194528875\ldots~ \\
6 & ~~ & ~0.01715448442\ldots~& ~~ & ~0.07755851308\ldots~ \\
7 & ~~ & ~0.01213788668\ldots~& ~~ & ~0.06708407871\ldots~ \\
8 & ~~ & ~0.009053001692\ldots~& ~~ & ~0.05911229290\ldots~ \\
9 & ~~ & ~0.007016561297\ldots~& ~~ & ~0.05283957845\ldots~ \\
10 & ~~ & ~0.005600098814\ldots~ & ~~ & ~0.04777380565\ldots~\\
11& ~~ & ~0.004574393818\ldots~ & ~~ & ~0.04359650569\ldots~\\
12& ~~ & ~0.003807469357\ldots~ & ~~ & ~0.04009237283\ldots~\\
13& ~~ & ~0.003218849539\ldots~ & ~~ & ~0.03711056811\ldots~\\
\hline
\end{tabular}
\label{bond-fcc}
\end{table}

In Table \ref{site-kag}, we present evaluations of the
best lower bound (\ref{eta-2-1}) on the percolation threshold
$p_c$ for site percolation on the non-Bravais lattices $\mbox{Dia}_d$ and $\mbox{Kag}_d$ up
through dimension 13 and compare them to the corresponding simulation
data when available. The results for bond percolation on these
lattices are given in Table \ref{bond-kag}. Again, it can be seen
that (\ref{eta-2-1}) already provides a tight bound on the
numerical estimates of $p_c$ in relatively low dimensions (e.g.,
$d=5$ and 6). Again, as in the cases of $D_d$ and $A_d^*$ lattices
described above, it is reasonable to expect that
our tightest lower bound (\ref{eta-2-1}) provides sharp estimates of $p_c$
for $d \ge 7$, especially in high dimensions.
These results are particularly useful in the absence of numerical evaluations 
of $p_c$ for such higher dimensions.

\begin{table}[!htp]
\caption{Comparison of numerical estimates of the site percolation
thresholds on the $\mbox{Kag}_d$ and $\mbox{Dia}_d$ lattices to
the corresponding best lower bounds on $p_c$, denoted by $p_c^L$.
Simulation results for $d=3-6$ for the $\mbox{Kag}_d$ lattice are
taken from Ref.  \onlinecite{Va98a}. Simulation results for $d=2$
and $d=3-6$ for the $\mbox{Dia}_d$ lattice are taken from Refs.
\onlinecite{Zi09} and \onlinecite{Va98b}, respectively. Note that
in the case $\mbox{Kag}_2$, $p_c=1-2\sin(\pi/18) = 0.6527036446\ldots$ is an exact
result \cite{Sy64}.}
\begin{tabular}{c|c|c|c|c }
\hline
&  \multicolumn{2}{|c|}{$\mbox{Kag}_d$} & \multicolumn{2}{|c}{$\mbox{Dia}_d$} \\
~Dimension~& ~$p_c$~ & ~$p_c^L$ from Eq.~(\ref{eta-2-1})& ~$p_c$~ & ~$p_c^L$ from Eq.~(\ref{eta-2-1}) \\
\hline
2 & ~0.6527036446\ldots~ & ~0.5000000000\ldots~ & ~0.6970413~& ~0.5000000000\ldots~\\
3 & ~0.3895~             & ~0.3333333333\ldots~ & ~0.4301~& ~0.3333333333\ldots~\\
4 & ~0.2715 ~           & ~ 0.2500000000\ldots~ & ~0.2978~ & ~0.2500000000\ldots~ \\
5 & ~0.2084 ~           & ~ 0.2000000000\ldots~& ~0.2252~ & ~0.2000000000\ldots~ \\
6 & ~0.1677~           & ~  0.1666666666\ldots~& ~0.1799~ & ~0.1666666666\ldots~ \\
7 & ~~                   & ~0.1428571429\ldots~& ~~ & ~0.1428571429\ldots~ \\
8 & ~~                   & ~0.1250000000\ldots~& ~~ & ~0.1250000000\ldots~ \\
9 & ~~                   & ~0.1111111111\ldots~& ~~ & ~0.1111111111\ldots~ \\
10 & ~~                  & ~0.1000000000\ldots~ & ~~ & ~0.1000000000\ldots~\\
11& ~~ & ~0.09090909090\ldots~ & ~~ & ~0.09090909090\ldots~\\
12& ~~ & ~0.08333333333\ldots~ & ~~ & ~0.08333333333\ldots~\\
13& ~~ & ~0.07692307692\ldots~ & ~~ & ~0.07692307692\ldots~\\
\hline
\end{tabular}
\label{site-kag}
\end{table}

\begin{table}[!htp]
\caption{Comparison of numerical estimates of the bond percolation
thresholds on the $\mbox{Kag}_d$ and $\mbox{Dia}_d$ lattices to
the corresponding best lower bounds on $p_c$. Simulation results
for $d=3-5$ for the $\mbox{Kag}_d$ lattice are taken from Ref.
\onlinecite{Va98b}. Simulation results for $d=2$ and $d=3-6$ for
the $\mbox{Dia}_d$ lattice are taken from Refs. \onlinecite{Di10}
and \onlinecite{Va98a}, respectively.  Note that in the case
$\mbox{Dia}_2$, $p_c=1-2\sin(\pi/18) = 0.6527036446\ldots$ is an exact result
\cite{Sy64}.}
\begin{tabular}{c|c|c|c|c }
\hline
&  \multicolumn{2}{|c|}{$\mbox{Kag}_d$} & \multicolumn{2}{|c}{$\mbox{Dia}_d$} \\
~Dimension~& ~$p_c$~ & ~$p_c^L$ from Eq.~(\ref{eta-2-1}) & ~$p_c$~ & ~$p_c^L$ from Eq.~(\ref{eta-2-1})\\
\hline
2 & ~0.524404978~ & ~0.4000000000\ldots~ & ~0.6527036446~& ~0.5000000000\ldots~  \\
3 & ~0.2709~      & ~0.2395833333\ldots~ & ~0.3895~     & ~0.3333333333\ldots~ \\
4 & ~0.177~       & ~0.1660649819\ldots~&   ~0.2715~    & ~0.2500000000\ldots~ \\
5 & ~0.130~       & ~0.1260229133\ldots~ & ~0.2084~     & ~0.2000000000\ldots~\\
6 & ~~           & ~0.1012216405\ldots~ & ~0.1677~     & ~0.1666666666\ldots~\\
7 & ~~          & ~0.08445595855\ldots~ & ~~     & ~0.1428571429\ldots~\\
8 & ~~          & ~0.07240119562\ldots~ & ~~     & ~0.1250000000\ldots~\\
9 & ~~          & ~0.06333107956\ldots~ & ~~     & ~0.1111111111\ldots~\\
10 & ~~         & ~0.05626598465\ldots~ & ~~     & ~0.1000000000\ldots~\\
11& ~~          & ~0.05061061531\ldots~ & ~~     & ~0.09090909090\ldots~ \\
12& ~~          & ~0.04598313360\ldots~ & ~~     & ~0.08333333333\ldots~ \\
13& ~~          & ~0.04212768882\ldots~ & ~~     & ~0.07692307692\ldots~ \\
\hline
\end{tabular}
\label{bond-kag}
\end{table}

\begin{figure}[htp]
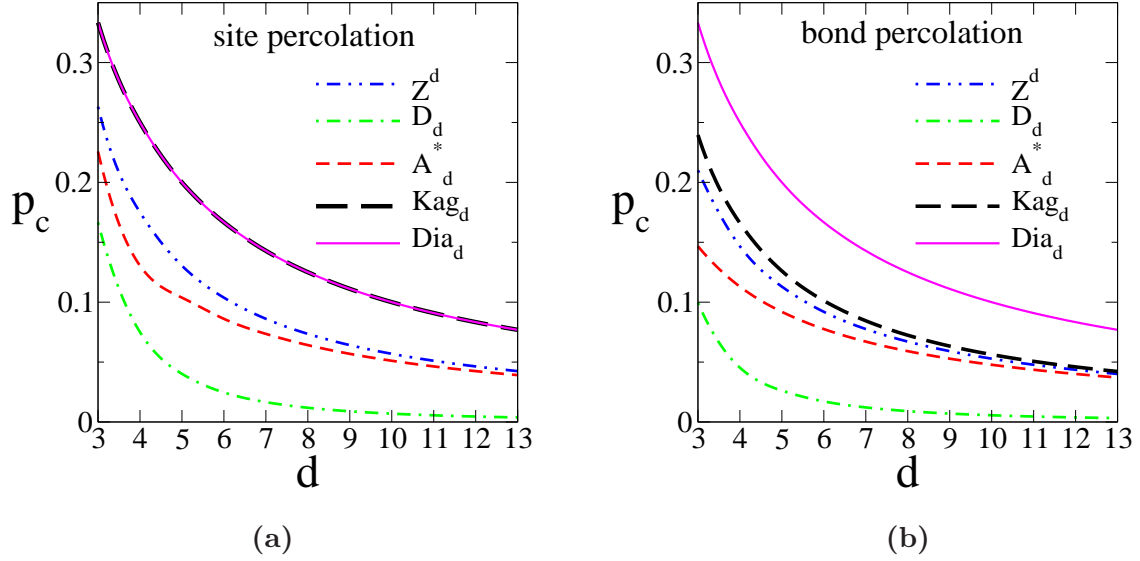

\begin{center}
\includegraphics[height=6.5cm,keepaspectratio]{2_1_site.eps}
\hspace{0.35in}\includegraphics[height=6.5cm,keepaspectratio]{2_1_bond.eps}\\
\hspace{0.2in}\mbox{\bf (a)} \hspace{3in} \mbox{\bf (b)}
\end{center}
\caption{Percolation threshold $p_c$ versus dimension $d$ for site and bond
percolation on the $d$-dimensional lattices $\mathbb{Z}^d$,
$D_d$, $A^*_d$, $\mbox{Dia}_d$ and $\mbox{Kag}_d$ as obtained from
the lower bound (\ref{eta-2-1}). (a) Site percolation. Note that lower bounds  for
$\mbox{Dia}_d$ and $\mbox{Kag}_d$ are identical. (b) Bond percolation.} \label{fig_site}
\end{figure}

It is clear that the tightest lower bound (\ref{eta-2-1}) on $p_c$
is accurate enough to enable us to compare the relative
trends of the thresholds for different lattices
in any fixed dimension $d$. Figure \ref{fig_site}  shows
the best lower bound (\ref{eta-2-1}) on $p_c$ for site and bond percolation on the 
five different $d$-dimensional
lattices $\mathbb{Z}^d$, $D_d$, $A^*_d$, $\mbox{Dia}_d$ and
$\mbox{Kag}_d$. For any fixed dimension, we see, not surprisingly, that the threshold on
$D_d$ is minimized among all of these lattices for either site or bond percolation due to the 
fact that it possesses the largest coordination number
$z_{D_d}$. Similarly, the local coordination structure
of the other lattices explains the trends in their relative threshold values.
Observe that in the case of site percolation, the lower bound on $p_c$ for $\mbox{Dia}_d$ is identical to that for
$\mbox{Kag}_d$, since the two percolation problems are exactly equivalent to one another (see Sec.IV.B).

\begin{figure}[htp]
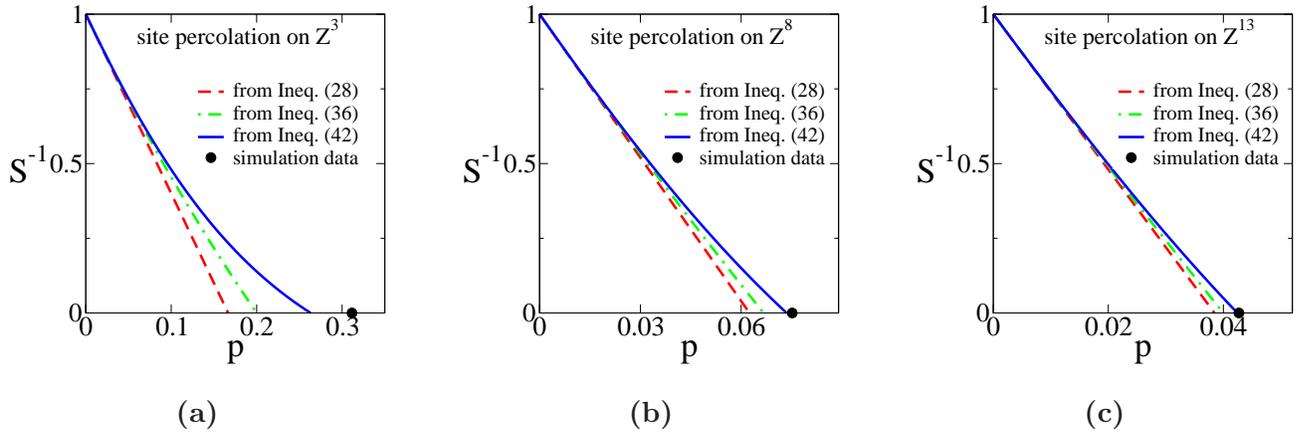

\begin{center}
$\begin{array}{c@{\hspace{1.0cm}}c@{\hspace{1.0cm}}c}\\
\includegraphics[height=4.8cm,keepaspectratio]{Zn_S.d_3.eps} &
\includegraphics[height=4.8cm,keepaspectratio]{Zn_S.d_8.eps} &
\includegraphics[height=4.8cm,keepaspectratio]{Zn_S.d_13.eps} \\
\mbox{\bf (a)} & \mbox{\bf (b)} & \mbox{\bf (c)}
\end{array}$
\end{center}
\caption{The lower bounds on the inverse of the mean cluster
number $S^{-1}$ versus $p$ for site percolation on hypercubic
lattice $\mathbb{Z}^d$ for (a) $d=3$, (b) $d=8$ and (c) $d=13$ as
obtained from the upper bound on $S$ (\ref{0-1}), (\ref{1-1}) and
(\ref{2-1}). Included in this figure are the percolation
thresholds (black circles) obtained from the accurate
numerical study of Ref.  \onlinecite{Ne00}.} \label{fig_Zd}
\end{figure}

Figure \ref{fig_Zd}  shows the lower bounds on the inverse of
the mean cluster number $S^{-1}$ as a function of $p$ as obtained
from the upper bounds on $S$ (\ref{0-1}), (\ref{1-1}) and
(\ref{2-1}) for site percolation on the hypercubic lattice
$\mathbb{Z}^d$ for $d=3, 8$ and 13. The zero of $S^{-1}$ gives
the threshold  and we also include in Fig. \ref{fig_Zd} the associated numerical
estimates of $p_c$. These plots clearly illustrate
that the lower bounds on $S^{-1}$ become
increasingly more accurate as the space dimension increases.
This is not surprising, since all of these lower bounds become
asymptotically exact as the space dimension becomes large.
The best lower bound, as obtained from (\ref{2-1}), gives a highly accurate estimate
of the inverse mean cluster size already for $d=8$
and essentially should coincide with the exact result as evidenced
by the very near proximity of the zero of the lower bound
with the numerically estimated threshold $p_c$.

\section{Conclusions and Discussion}
\label{discuss}

We have shown that $[0,1]$, $[1,1]$ and $[2,1]$ Pad{\'e } approximants
of the mean cluster number $S$ for site and bond percolation on general
$d$-dimensional lattices are upper bounds on this quantity in any Euclidean dimension $d$.
These results immediately lead to lower bounds on the threshold $p_c$.
We obtain explicit bounds on $p_c$ for several types of lattices:
$d$-dimensional generalizations of the simple-cubic, body-centered-cubic
and face-centered-cubic Bravais lattices as well as the $d$-dimensional
generalizations of the diamond and kagom{\'e } (or pyrochlore) non-Bravais
lattices. We have calculated the lower bounds for these lattice
lattices and compared them to the available numerical estimates of $p_c$. 
The lower bounds on $p_c$ obtained from 
$[1,1]$ and $[2,1]$ Pad{\'e } approximants become asymptotically exact
in the high-$d$ limit. The best lower bound, obtained from the [2, 1] Pad{\'e} approximant, 
is relatively tight for $3 \le d \le 5$ and generally provides excellent estimates 
of $p_c$ for $d \ge 6$. While the [0,1] estimate of $p_c$ was proven to be a lower bound here,
rigorous proofs of that the [1,1] and [2,1] estimates are indeed
lower bounds will be reserved for a future publication. However, we have presented very strong evidence
that the latter are indeed lower bounds for the class
of $d$-dimensional lattices considered in this paper.

We have seen in Sec. V that the  estimate of $p_c$ obtained 
from the best lower bound (\ref{eta-2-1})
for both site and bond percolation on $D_d$ converges to the corresponding
numerical estimates in relatively low dimensions most rapidly among all of the five $d$-dimensional
lattices that we have studied in this paper. 
This is due to the highly connected nature of $D_d$, e.g., it possesses the largest coordination number $z_{D_d}$ among all of 
the five lattices studied here. In addition, we have shown 
that the asymptotic expansions of the lower-bound estimates are 
exact through at least $1/d^3$ and $1/d^4$, respectively for site and bond percolation 
on $D_d$, and therefore more accurate than the corresponding asymptotic
expansions for the other lattices. This observation is consistent with the principle that 
high-dimensional results encode information about percolation behavior in low dimensions,
as is also the case in continuum percolation \cite{To12a,To12c,To13a}.

Among all of the ten percolation problems that we considered in the paper, 
the only case in which the high-$d$ limit of the threshold $p_c$ does not 
agree with the corresponding Bethe approximation (\ref{B1}) is for site percolation on 
the $d$-dimensional kagom{\'e} lattice $\mbox{Kag}_d$. The usual 
arguments explaining the tendency of a lattice 
to behave like an infinite Bethe tree \cite{Fi61a} apply in all of the other nine 
cases. For example, consider bond percolation on $\mbox{Dia}_d$, which 
gives $p_c \sim 1/d$ (i.e., the Bethe approximation). This is the only specific instance
in which a bond percolation problem can be exactly mapped to a site 
percolation problem, namely, that on the kagom{\'e} lattice $\mbox{Kag}_d$. 
Therefore, while the coordination number of the latter $z_{\mbox{\tiny Kag}_d} = 2d$, 
the threshold $p_c$ for site percolation on $\mbox{Kag}_d$ must, in any dimension, agree 
with that for bond percolation on $\mbox{Dia}_d$ and hence $p_c$ must 
tend to $1/d$ [not $1/z_{\mbox{\tiny Kag}_d} = 1/(2d)$] in the high-$d$ limit.

It was once hypothesized that the percolation threshold of a lattice corresponded to 
the radius of convergence of the series expansion
for $S$ \cite{Do61}. This hypothesis rested on the assumption that $S$ had no singularities
on the positive real axis for $p$ less than the critical value, i.e.,
the coefficients $S_2,S_3,\ldots$ were all positive. It was shown that at sufficiently high
order (e.g., 19th-order), the coefficients are sometimes negative for $d=2$. This implies that
the critical concentration does not correspond to the radius of convergence
of the series expansion for $S$ for $d=2$, strongly suggesting that
there is a closer singularity on the negative real axis \cite{Sy73}.

In analogy with the continuum percolation results of Ref. \onlinecite{To12a},
our present results offer evidence that in sufficiently high dimensions,
the radius of convergence of (\ref{S5}) for Bernoulli lattice percolation corresponds to $p_c$. 
The fact that the putative lower bound on $p_c$ [c.f. Eq. (\ref{eta-1-1})]
obtained from the [1, 1] Pad{\'e} approximant of $S(p)$ [c.f., Eq. (\ref{1-1})] 
is asymptotically exact through second order terms implies that 
Eq. (\ref{1-1}) is also asymptotically exact, i.e., 
\begin{equation}
\label{S_pole}
S(p) \sim \frac{1}{1-\frac{S_3}{S_2}p}, \qquad d \rightarrow \infty,
\end{equation}
with critical exponent $\gamma=1$ (cf. (\ref{Sasymp})], as expected.
This in turn implies that the radius of convergence in the high-dimensional
limit corresponds to the percolation threshold $p_c=S_2/S_3$ because all of the coefficients of 
of the resulting expansion of $S(p)$ [cf. Eq. (\ref{S5})] are all positive.
Recall that for $d=1$, $S$ is given by (\ref{pole-1}), and hence all of the coefficients
$S_m$ are positive.
Thus, it appears that the closest singularities for
the occupation probability $p$ expansion of $S(p)$ shifts from the positive real axis to
the negative real axis in going from one to two dimensions, 
remain on the negative real axis for sufficiently low dimensions
$d \ge 3$, and eventually move back to the positive real axis for sufficiently large $d$.


\section*{Acknowledgements}
\vspace{-0.2in}

We are grateful to Michael Aizenman for useful discussions.
This work was supported by the Materials Research Science
and Engineering Center  Program of the National
Science Foundation under Grant No. DMR-0820341.
S.T. gratefully acknowledges the
support of a Simons Fellowship in Theoretical Physics, which
has made his sabbatical leave this entire academic year
possible.

\appendix

\renewcommand{\theequation}{A-\arabic{equation}} 
\setcounter{equation}{0}  

\section*{Appendix A: Analytical Determination of Cluster Statistics
for $d$-Dimensional Lattices}

In this appendix, we describe the algorithm that we have used to 
obtain analytical expressions for the coefficients $S_2(d)$, $S_3(d)$ 
and $S_4(d)$ in the series expansion of the mean cluster number $S$ in 
powers of the site (bond) occupation probability $p$ for any lattice 
in {\it high dimensions} presented in Sec.  \ref{Sk}. As discussed in Sec. II, $S$ can expressed in 
terms of the cluster-size distribution function $n_k$ [c.f. Eq. (\ref{S4})]. 
Therefore, it is sufficient for us to determine the expressions of $n_k$ 
[c.f. Eq. (\ref{sum2})], from which the series expansion of $S$ can be obtained 
in any specific $d$.  
The general $d$-dimensional coefficient $S_k(d)$
can then be determined using the fact
that it is a polynomial in $d$, i.e.,
\begin{equation}
S_k(d) = \sum_{n=1} \kappa_n d^n.
\label{A_Sk}
\end{equation}
The coefficients $\kappa_n$ are determined by solving a set of 
linear equations in the first several dimensions (e.g.,
$2\le d\le 5$) such that they satisfy the explicitly known
forms for $S_k$ in these relatively low dimensions.

\begin{figure}[bthp]
\begin{center}
$\begin{array}{c@{\hspace{1.5cm}}c}\\
\includegraphics[height=2.8cm,keepaspectratio]{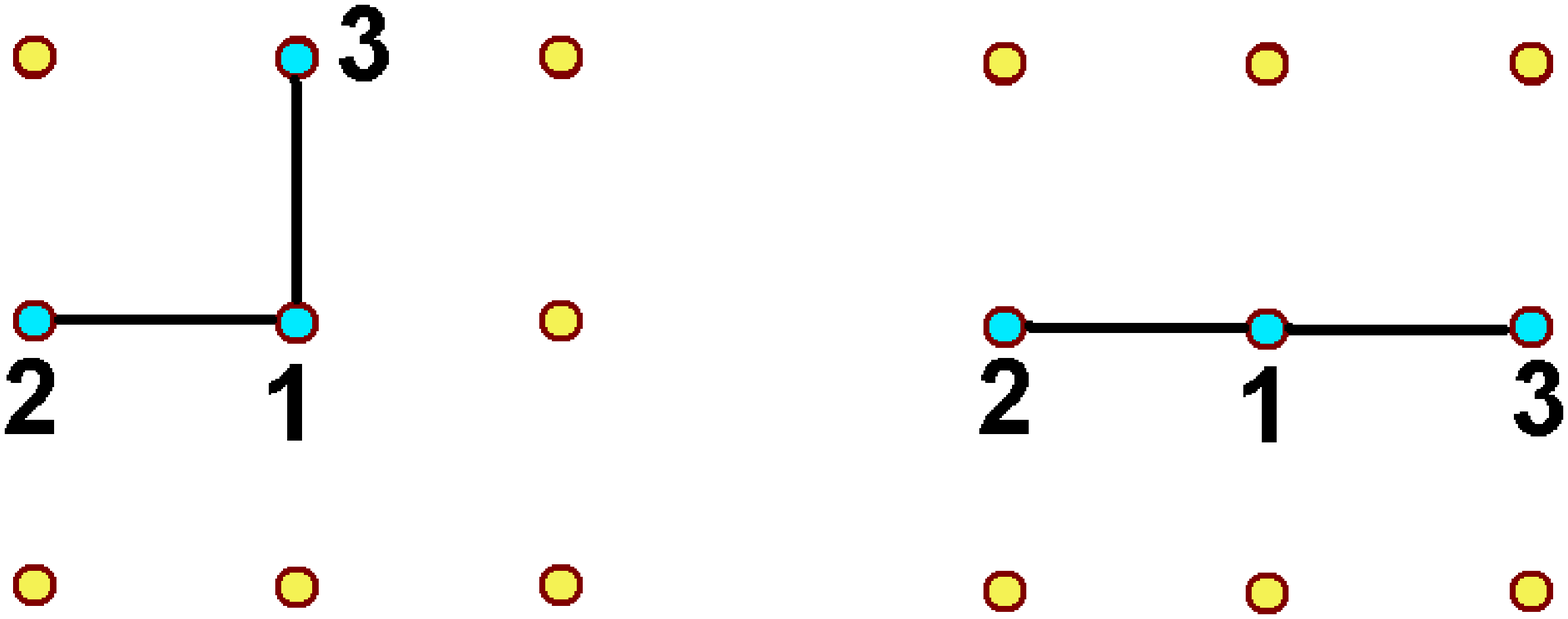} &
\includegraphics[height=2.8cm,keepaspectratio]{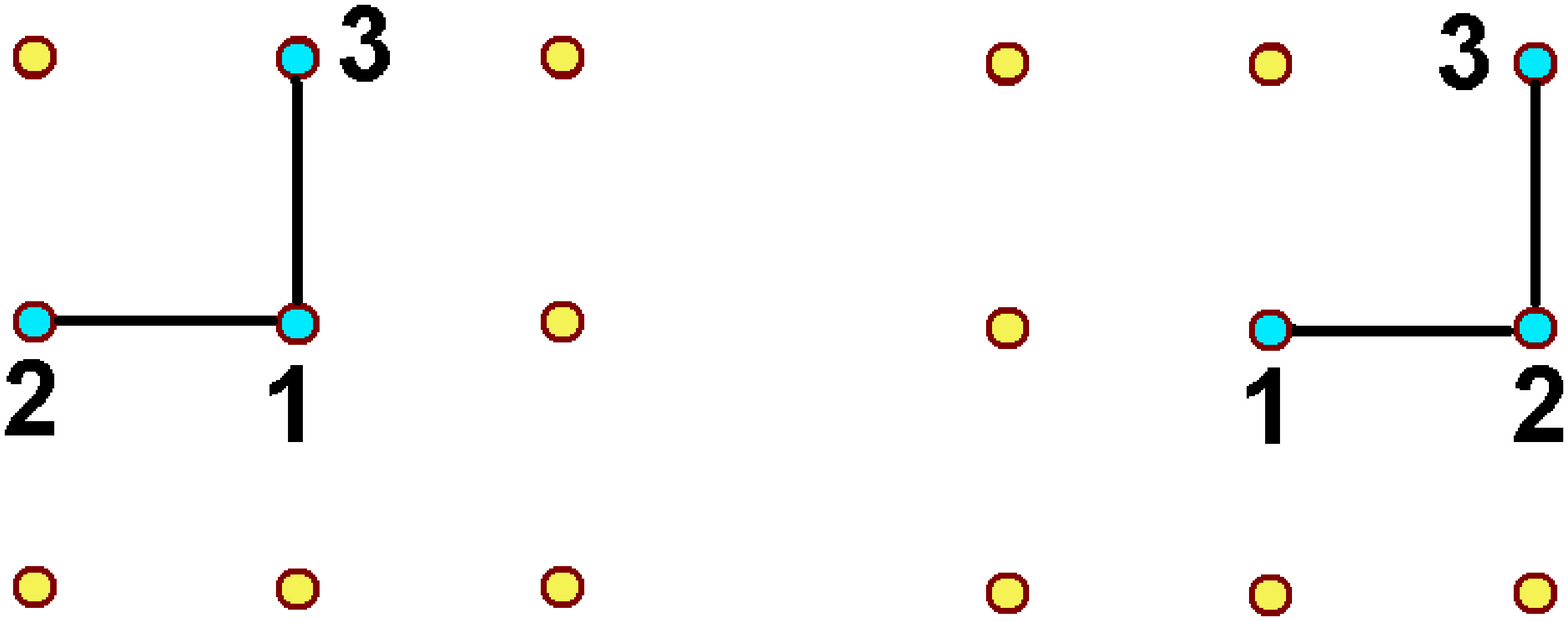} \\
\mbox{\bf (a)} & \mbox{\bf (b)}
\end{array}$
\end{center}
\caption{(color online). Two pairs of distinct trimer configurations associated with site percolation 
on the square lattice $\mathbb{Z}^2$. Note that the numbers indicate cluster site labels 
rather than ordered labels of the sites of $\mathbb{Z}^2$. (a) Two distinct trimer configurations that cannot be mapped 
to one another by any translation or rotation. (b) Two distinct trimer configurations that 
can be mapped to one another by a simple translation.} \label{fig_square}
\end{figure}

Our algorithm enables us to obtain analytically the polynomials  $n_k$ by 
directly enumerating all of the distinct $k$-mer configurations associated with a site (bond) located 
at, without loss of any generality, some chosen origin.
We note that two $k$-mer configurations are distinct if they contain one or more distinct sites (bonds);
see Fig. \ref{fig_square} for simple examples. 
To the best of our knowledge, such an algorithm has not been applied before to 
obtain explicit expressions for the $n_k$'s. 
Our algorithm works as follows: For a given $d$-dimensional lattice, the vectors connecting a site (bond) to all of 
its nearest neighbors are determined. All of the $k$-mer configurations associated with a selected 
site (bond) are then generated. Specifically, a $k$-mer configuration is generated from a $(k-1)$-mer configuration
($k \ge 2$) by adding a site (bond) that is a nearest neighbor of one of the sites 
(bonds) in the $(k-1)$-mer configuration. The total number of $k$-mer configurations for a site 
($k-1$-mer configurations for a bond) generated 
in this way is $(k-1)!z_{\Lambda}^{(k-1)}$, where $z_{\Lambda}$ is the 
coordination number of the given lattice $\Lambda$. Although in principle 
this algorithm can be employed to obtain cluster statistics for arbitrary $k$, 
we are only interested in the cases where $1 \le k\le 4$ here, but for any dimension $d$.

The $k$-mer configurations are then compared to one another to obtain the set of distinct $k$-mer configurations. 
For site percolation, we find that the set of vector displacements between any two 
sites is sufficient to distinguish a pair of $k$-mer configurations. For bond percolation, a $k$-mer 
contains $k$ bonds and $\gamma$ associated sites (e.g., $\gamma=k+1$ is the $k$-mer does not contain 
closed loops). The latter is simply a $\gamma$-mer in the site context. 
A $k$-mer configuration containing $k$ bonds can be mapped into a configuration of $k$ points
by placing the points at the midpoints of any bond.
Note that these midpoints are not sites of the given lattice, but rather
a new ``site" decoration of the lattice. The vector-displacement sets for both the 
$\gamma$-mer configurations of the sites and the configuration the mapped $k$ points are required to 
distinguish two $k$-mer configurations of bonds. In particular, a displacement matrix ${\bf M}^{\alpha\beta}$ 
is used to distinguish a pair of $k$-mer configurations, $\alpha$ and $\beta$. The components of 
the matrix $M^{{\alpha\beta}}_{ij}$ are the vector displacements between two 
sites (points) $i$ and $j$, one in each $k$-mer configuration (point configuration). Two $k$-mer configurations are 
identical if every row of ${\bf M}^{\alpha\beta}$ has at least one component that is a zero vector.

Figure \ref{fig_square} shows two simple examples of how the vector-displacement 
matrix ${\bf M}^{\alpha\beta}$ can be applied to distinguish a pair of 
trimer configurations (i.e., clusters of three sites) for site percolation on the 
square lattice $\mathbb{Z}^2$. Figure \ref{fig_square}(a) shows 
two distinct trimer configurations that cannot be mapped to one another by any translation or rotation.
The associated displacement matrix is given by 
\begin{equation}
{\bf M}^{\alpha\beta} = \left [{\begin{array}{ccc} 
~(0, 0)~ & ~(-1, 0)~ & ~(1, 0)~ \\
~(1, 0)~ & ~(0, 0)~ & ~(2, 0)~ \\
~(0, -1)~ & ~(-1, -1)~ & ~(1, -1)~ \\
\end{array}}\right ],
\label{A_M1}
\end{equation}
which does not satisfy the condition that every row has at least one zero vector.
Note that we have set the distance between two nearest neighbor sites to be unity 
and the entry $M^{{\alpha\beta}}_{11}$ is always zero since it is associated with the 
common origin for any $k$-mer configuration. Figure \ref{fig_square}(b) shows 
two distinct trimer configurations that can be mapped to one another a simple translation.
The associated displacement matrix is given by 
\begin{equation}
{\bf M}^{\alpha\beta} = \left [{\begin{array}{ccc} 
~(0, 0)~ & ~(1, 0)~ & ~(1, 1)~ \\
~(1, 0)~ & ~(2, 0)~ & ~(2, 0)~ \\
~(0, -1)~ & ~(1, -1)~ & ~(1, 0)~ \\
\end{array}}\right ].
\label{A_M2}
\end{equation}
While the matrix does not have zero vectors in every row, 
the vector $(1, 0)$ is contained in every row, which is the translation 
vector that maps the two trimer configurations to one another. It is clear 
that if the translation vector is a zero vector, the two trimer configurations
are then identical. 


Finally, for each distinct $k$-mer configuration, the number of vacate sites
(bonds) that are nearest neighbors of the sites (bonds) in the
$k$-mer configuration is determined, which gives the value of the associated
$m$ (i.e., the exponent associated with $1-p$ term in Eq. (\ref{sum2}). 
Since distinct $k$-mer configurations that can be obtained from one another
by simple rotation or translation have the same vacancy
configuration, they contribute identical terms to the polynomials
for $n_k$. The total number of such $k$-mers gives the value of
the associated coefficient $g_{km}$.

For five of the ten percolation problems considered in this paper, the expressions for the $n_k$'s can
be explicitly written as a function of dimensionality $d$, which are provided here.
Explicit expressions for  $n_1$, $n_2$,
$n_3$ and $n_4$  in dimensions 2 to 5 for all of the ten percolation problems
are provided in the Supplemental Material.

For site percolation on hypercubic lattice
$\mathbb{Z}^d$, the $n_k$'s  are given by
\begin{equation}
\begin{array}{c}
n_1 = p(1-p)^{2d}, \\
n_2 = dp^2(1-p)^{4d-2}, \\
n_3 = 2d(d-1)p^3(1-p)^{6d-5} + 2d(d-1)p^3(1-p)^{6d-4}, \\
n_4 = \frac{4}{3}d(d-1)(d-2)p^4(1-p)^{8d-9}+\frac{1}{2}d(d-1)(8d-7)p^4(1-p)^{8d-8}+ \\
+4d(d-1)p^4(1-p)^{8d-7}p^4+dp^4(1-p)^{8d-6}.
\end{array}
\end{equation}
For bond percolation on hypercubic lattice
$\mathbb{Z}^d$, the $n_k$'s are given by
\begin{equation}
\begin{array}{c}
n_1= p(1-p)^{4d-2}, \\
n_2= (2d-1)p^2(1-p)^{6d-4}, \\
n_3= 2d(d-1)p^3(1-p)^{8d-7} + \frac{1}{3}(16d^2-24d+11)p^3(1-p)^{8d-6}, \\
n_4 =\frac{1}{2}(d-1)p^4(1-p)^{8d-8}+16(d-1)^2p^4(1-p)^{10d-9}+ \\
+\frac{1}{12}(200d^3-552d^2+574d-210)p^4(1-p)^{10d-8}.\\
\end{array}
\end{equation}

For site percolation on $d$-dimensional diamond
lattice $\mbox{Dia}_d$, the $n_k$'s are given by
\begin{equation}
\begin{array}{c}
n_1= p(1-p)^{d+1}, \\
n_2= \frac{1}{2}(d+1)p^2(1-p)^{2d}, \\
n_3= \frac{1}{2}d(d+1)p^3(1-p)^{3d-1}, \\
n_4 =\frac{1}{6}d(3d^2-d+4)p^4(1-p)^{4d-2}.\\
\end{array}
\end{equation}
For bond percolation on $d$-dimensional diamond
lattice $\mbox{Dia}_d$, the $n_k$'s are given by
\begin{equation}
\begin{array}{c}
n_1= p(1-p)^{2d}, \\
n_2= dp^2(1-p)^{3d-1}, \\
n_3= \frac{1}{3}d(4d-1)p^3(1-p)^{4d-2}, \\
n_4 =\frac{1}{12}d(5d-1)(5d-2)p^4(1-p)^{5d-3}.\\
\end{array}
\end{equation}

For site percolation on $d$-dimensional kagom{\'e}
lattice $\mbox{Kag}_d$, the $n_k$'s are given by
\begin{equation}
\begin{array}{c}
n_1= p(1-p)^{2d}, \\
n_2= dp^2(1-p)^{3d-1}, \\
n_3= \frac{1}{3}d(4d-1)p^3(1-p)^{4d-2}, \\
n_4 =\frac{1}{12}d(5d-1)(5d-2)p^4(1-p)^{5d-3}.\\
\end{array}
\end{equation}
Note that these expressions of $n_k$'s are identical to
those for bond percolation on $\mbox{Dia}_d$.

\section*{Appendix B: Explicit Calculation of $S_3$ Using Eq. (\ref{s3}) For Site Percolation in $A^*_2$}


In this appendix, we explicitly calculate $S_3$ using Eq.
(\ref{s3}) for site percolation on the triangular lattice (i.e.,
$A^*_2$) as an instructive illustration of how to obtain $S_k$
directly from the connectivity function $f$. Since the function
$f$ is only nonzero for a pair of bonds that are nearest neighbors
of one another [see Eq. (\ref{s3})], it is clear that only when
site $k$ is not a nearest neighbor of site 1 and when site $j$ is
a mutual nearest neighbor of sites 1 and $k$, the product in the
double sum has nonzero value (i.e., unity). This also suggests
that site $k$ can be at most two bonds away from site 1, otherwise
it cannot share a common nearest neighbor with site 1.

\begin{figure}[bthp]
\begin{center}
$\begin{array}{c@{\hspace{1.5cm}}c}\\
\includegraphics[height=4.0cm,keepaspectratio]{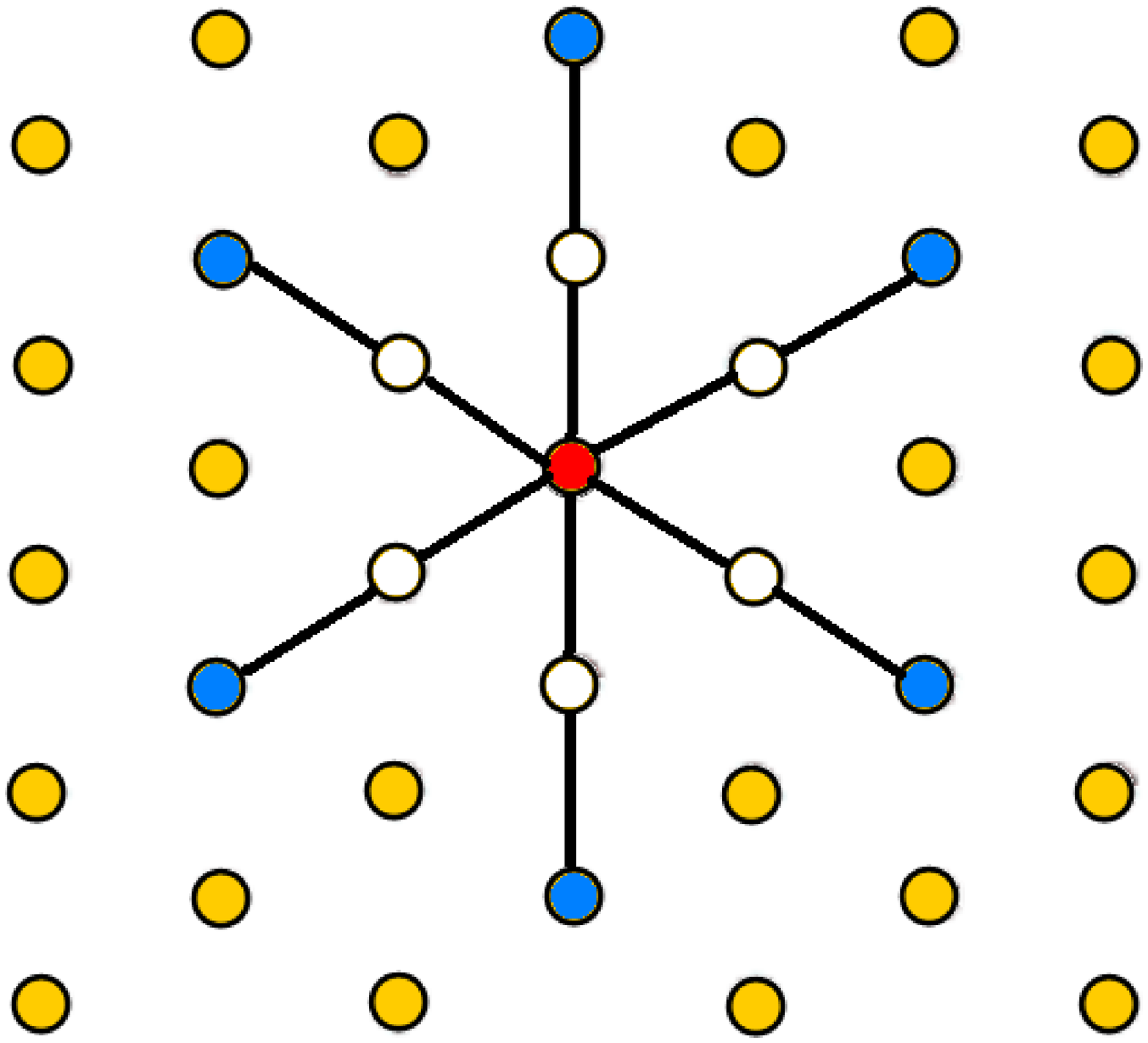} &
\includegraphics[height=4.0cm,keepaspectratio]{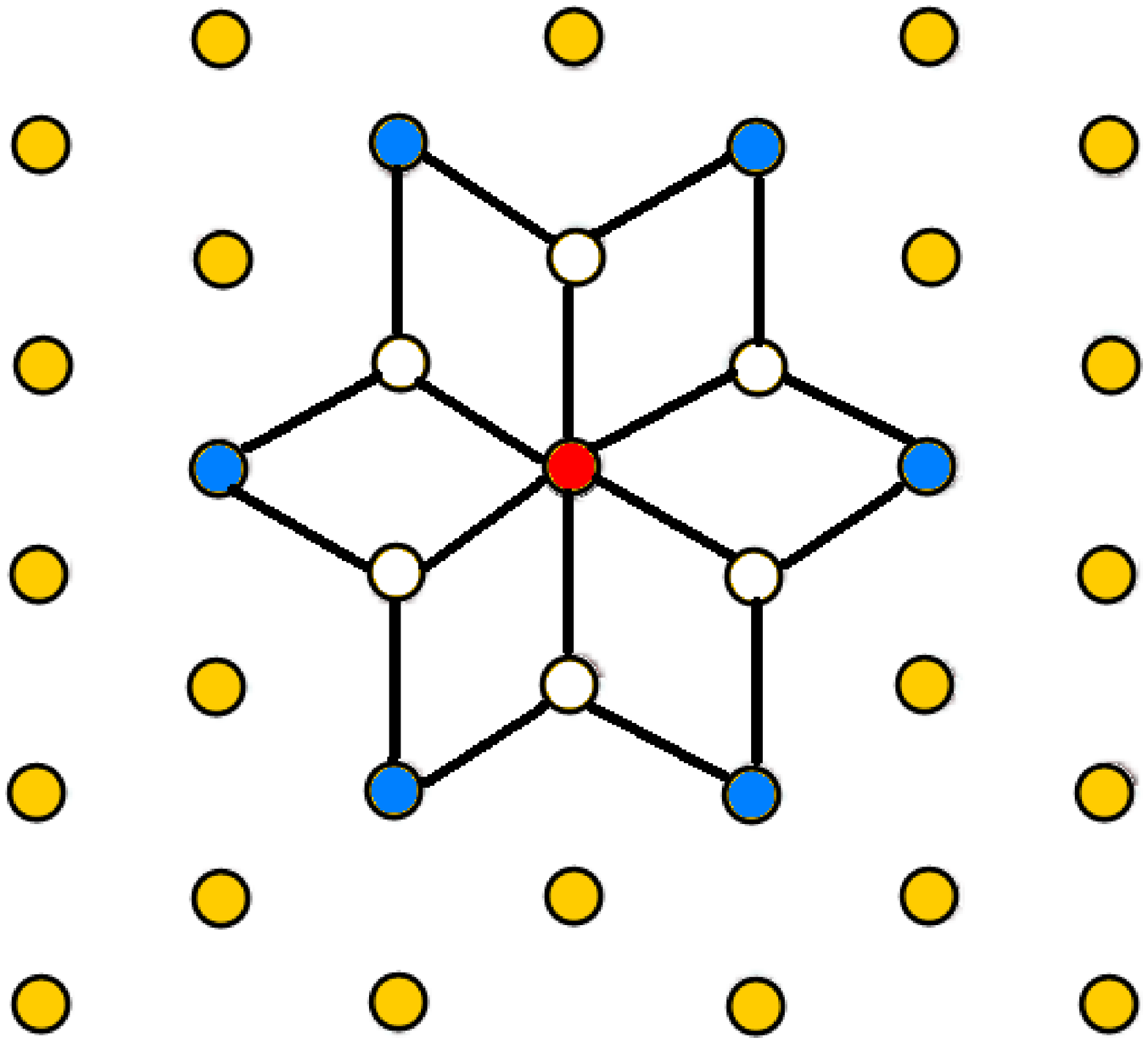} \\
\mbox{\bf (a)} & \mbox{\bf (b)}
\end{array}$
\end{center}
\caption{(color online). Three-site clusters (3-mers) of the
triangular lattice that contribute to the coefficient $S_4$. (a)
The lineal configuration in which sites 1 (red or dark
gray) and $k$ (blue or light gray) are connected by a single
common nearest neighbor $j$ (empty circles). (b) The non-lineal
configurations in which sites 1 (red or dark gray) and $k$ (blue
or light gray) can be connected by two common nearest neighbors
$j$ (empty circles).} \label{fig_tri}
\end{figure}

Figure \ref{fig_tri} shows two configurations of sites 1 and $k$
that contribute to Eq. (A-1). In the first configuration (Fig.
\ref{fig_tri}a), sites 1 and $k$ form a straight line and can be
connected by the common nearest neighbor $j$ in between. Due to
the symmetry of the lattice, there are 6 such lineal
configurations, each contributing 1 to $S_3$. In the second
configuration (Fig. \ref{fig_tri}b), each pair of sites 1 and $k$
can be connected by 2 common nearest neighbors $j$, which form a
folded line. Again, due to the symmetry of the lattice, there are
12 such non-lineal configurations, each contributing 1 to $S_3$.
Thus, we have
\begin{equation}
S_3 = 6\times 1\mbox{ (lineal configurations)}+12\times
1\mbox{ (non-lineal configurations)} = 18.
\end{equation}

We note that both the lineal and non-lineal configurations
are 3-site clusters. One might initially think that a simple counting of all
3-site clusters would lead to the same result. Although such a counting procedure would
lead to the correct result for some special cases, such as site percolation on the
square lattice, it is generally is not valid. For example, the
equilateral-triangle 3-site clusters do not contribute to $S_3$
here. This naive counting procedure would lead to an overestimation of
$S_3$.



\end{document}